\documentclass[fleqn,usenatbib]{mnras}

\usepackage{mathptmx}
\usepackage{txfonts}

\usepackage[T1]{fontenc}
\usepackage{ae,aecompl}

\usepackage{times}
\usepackage{amssymb}
\usepackage{amsmath}
\usepackage{color}
\usepackage{bm}
\usepackage{graphicx}
\usepackage{tabu}
\usepackage{epsfig}
\usepackage{longtable,tabu}
\usepackage{url}
\usepackage{ulem}
\usepackage{supertabular}
\usepackage{caption}
\usepackage[multidot]{grffile}
\usepackage{subcaption}
\usepackage[dvipsnames]{xcolor}

\def \herschel {\it {Herschel}}
\def \mum {$\rm \mu m$}
\def \lir {$L_{\rm IR}$}
\def \mstar {$M_{*}$}

\def \lco {$L^{'}_{\rm CO}$}
\def \lrad {$L_{\rm 1.4GHz}$}
\def \msun{$\rm M_{\odot}$}

\def \rhom{$\rho_{(M_{\rm mol})}$}

\title[Cosmic evolution of molecular gas mass density]
{Cosmic evolution of molecular gas mass density from an empirical relation between \lrad\ and \lco}

\author[G. Orellana-Gonz\'alez et al.]{%
G.~Orellana-Gonz\'alez,\(^{\! 1,2,3}\)\thanks{gustavo.orellana@uv.cl}
E.~Ibar,\(^{\! 1}\)
R.~Leiton,\(^{\! 1}\)
A.\,P.~Thomson,\(^{\! 4}\)
C.~Cheng,\(^{\! 5}\)
\and
R.\,J.~Ivison,\(^{\! 6}\)
R.~Herrera-Camus,\(^{\! 2}\)
H.~Messias,\(^{\! 7}\)
P.~Calder\'on-Castillo,\(^{\! 2}\)
\and
T.\,M.~Hughes\(^{8,1,9,10}\)
and L.~Leeuw\(^{11}\)
\vspace*{1mm}\\
$^{1}$Instituto de F\'isica y Astronom\'ia, Universidad de Valpara\'iso, Avda. Gran Breta\~na 1111, Valpara\'iso, Chile\\
$^{2}$Departamento de Astronom\'ia, Universidad de Concepci\'on, Casilla 160-C, 
Concepci\'on, Chile\\
$^{3}$Departamento de Matem\'atica y F\'isica Aplicadas, Universidad Cat\'olica de la Sant\'isima Concepci\'on, Alonso de Ribera 2850, Concepci\'on, Chile\\
$^{4}$Jodrell Bank Centre for Astrophysics, The University of Manchester, Oxford Road, Manchester, M13~9PL\\
$^{5}$Chinese Academy of Sciences South America Center for Astronomy, National Astronomical Observatories, CAS, Beijing 100101, China\\
Email: chengcheng@bao.ac.cn \\
$^{6}$European Southern Observatory, Karl-Schwarzschild-Strasse~2, D-85748 Garching, Germany\\
$^{7}$Astrof\'isica e Ci\^encias do Espa\c{c}o, Tapada da Ajuda - Edif\'icio Leste - 2o Piso
1349-018 Lisboa, Portugal\\
$^{8}$Chinese Academy of Sciences South America Center for Astronomy, China-Chile Joint Center for Astronomy,\\
Camino El Observatorio 1515,Las Condes, Santiago, Chile    \\
$^{9}$CAS Key Laboratory for Research in Galaxies and Cosmology, University of Science and Technology of China, Hefei 230026, China    \\
$^{10}$School of Astronomy and Space Science, University of Science and Technology of China, Hefei 230026, China \\  
$^{11}$College of Graduate Studies, University of South Africa, Theo van Wijk 9-52, Unisa Muckleneuk Campus, P.O.~Box 392,\\
UNISA, 0003, South Africa
}

\date{Accepted XXX. Received YYY; in original form ZZZ}

\pubyear{2020}

\begin{document}
\label{firstpage}
\pagerange{\pageref{firstpage}--\pageref{lastpage}}
\maketitle

\begin{abstract}
  Historically, GHz radio emission has been used extensively to
  characterize the star-formation activity in galaxies. In this work,
  we look for empirical relations amongst the radio luminosity, the
  infrared luminosity, and the CO-based molecular gas mass. We assemble
  a sample of 278 nearby galaxies with measurements of radio continuum
  and total infrared emission, and the $^{12}$CO ($J = 1$--0) emission
  line. We find a correlation between the radio continuum and the CO
  emission line (with a scatter of 0.36 dex), in a large sample
    of different kind of galaxies. Making use of this correlation, we
  explore the evolution of the molecular gas mass function and the
  cosmological molecular gas mass density in six redshift bins up
    to $z = 1.5$. These results agree with previous semi-analytic
  predictions and direct measurements: the cosmic molecular gas
    density increases up to $z=1.5$. In addition, we find a single
  plane across five orders of magnitude for the explored
    luminosities, with a scatter of 0.27 dex. These correlations are
  sufficiently robust to be used for samples where no CO measurements
  exist.
\end{abstract}

 


\section{Introduction}
\label{sec:intro}

Understanding galaxy evolution is a key goal of modern
  astrophysics. This can be tackled using a wide range of
  observational and theoretical prescriptions which involve complex
  physical processes for their interpretation (e.g.\
  \citealt{davies19}). In the cosmological context, a primary
  focus of this study is the evolution of the cosmic star-formation
history (CSFH, e.g.\ \citealt{tisley80,madau96}). Studies based on
large galaxy samples reveal that the cosmic star-formation rate (SFR)
density peaks at $z\sim2$--3 then declines steadily by
$\sim\,20\times$ through to the present time \citep{madau14}. The
physical processes involved are not yet clear, although some of the
main predictions are: the growth of the dark-matter halos (e.g.\
\citealt{behroozi13}), the depletion of molecular gas reservoirs in
galaxies (e.g.\ \citealt{kennicutt12,genzel15,tacconi18}) and changes
in the star-formation efficiency (e.g.\ \citealt{genzel10,daddi10}).

The molecular gas is the reservoir for the future star-formation (SF)
activity, so its census across cosmic time is fundamental to
understand CSFH. Historically, the main proxy to trace the molecular
gas mass is via the rotational low-$J$ ($J$ = 1--0) transitions of the
carbon monoxide (CO) molecule \citep{bolatto13}. Recently,
spectroscopic surveys undertaken with the Atacama Large
Millimeter/submillimeter Array (ALMA, e.g.\ ASPECS;
\citealt{walter16,decarli16}) have revealed that the molecular gas
mass density changes by a factor of 3--$10\times$ across $z=0$--2.
This result is in concordance with predictions from semi-analytic
models (e.g.\ \citealt{popping12,lagos18}). Nevertheless, when
CO-based studies are performed via pencil-beam surveys, they are: (a)
naturally affected by cosmic variance \citep{walter16}; (b) biased to
detect the most intensely star-forming galaxies (e.g.\
\citealt{bothwell13}) and (c) based on small galaxy samples (e.g.\
\citealt{daddi10,genzel10,genzel15}). These limitations are a
consequence of the considerable amount of observing time needed to
measure CO lines in galaxies beyond the local Universe.
    
An alternative but less direct method to measure the gas mass,
$M_{\rm gas}$, uses the dust that is concomitant with the molecular
gas. Several studies show that optically thin dust emission in the
Rayleigh-Jeans regime ($\sim 350$--1000\,\mum) is proportional to the
dust mass (e.g.\ \citealt{dunne11,clemens13,bianchi13}), where there
is a strong sensitivity on dust temperature \citep{scoville14}. 
  Diffuse,
  cold  dust ($T \leq 25$\,K)
  dominates the dust mass in galaxies, while warmer dust ($T
  \geq 30$\,K) usually dominates the dust luminosity
  \citep{devereux90,dunne01,draine07,clark15}. According to
\citet{scoville14}, the molecular gas mass can be obtained from a
single measurement of flux density made in a specific band on the
Rayleigh-Jeans tail, assuming a typical dust temperature between
20--45\,K. Adopting a dust-to-gas mass ratio, these authors then
determine the molecular gas mass. Using this method, recent studies
(e.g.\ \citealt{scoville16,hughes17a}) using small samples of massive,
nearby, star-forming, infrared-bright galaxies
($L_{\rm IR} > 10^{11}$\,L$_{\odot}$) found empirical relations
between the luminosity at 850\,\mum\ ($L_{850}$) and the total
$M_{\rm gas}$. The $L_{850}$---$M_{\rm gas}$ relation is tight
(scatter $\sim\,0.3$ dex), which makes it possible to obtain
$M_{\rm gas}$ in an efficient, precise manner for large samples of
galaxies.

Using a sample $10\times$ larger than that of \cite{scoville14},
however, \citet{orellana17} found that the relation between the
submillimetre (submm) emission and the gas mass determined by
Scovile's method shows a significant dispersion ($\sim\,1$ dex), as well
as increased scatter towards fainter luminosities. 
  Moreover, \citet{orellana17} found that the dust mass is a more
  accurate tracer of the total gas mass (atomic and molecular) than
  the molecular gas mass alone, in concordance with recent
  results obtained by \citet{casasola20} based on 436 nearby galaxies
  from the DustPedia Survey \citep{davies17}.

In this paper we construct new scaling relations using other tracers
of molecular gas which can provide more precise molecular gas mass
estimates with a lower dispersion.

To tackle this problem, we make use of three well-studied
  relations. First, the radio continuum--infrared (RC--IR)
correlation connects the non-thermal radio emission, typically at
1.4\,GHz, and the IR radiation coming from dust grains heated by
ultraviolet photons from young stellar populations, assuming that
supernovae remnants related to massive young stars are responsible for
the synchrotron radiation. The RC--IR correlation has been shown to be
valid over a wide range of star-forming galaxies (e.g.\
\citealt{bell03,ibar08,ivison10,smith14,liu15}) with little evolution
across the cosmic time \citep{ivison10,magnelli15}.

Second, the global Schmidt-Kennicutt (SK) relation connects the
formation of stars with the fuel to produce them, i.e.\ the molecular
gas; it can be expressed in terms \lir\ and \lco\ luminosities,
respectively (e.g.\ \citealt{kennicutt12}), where \lco\ is the
luminosity of the CO molecule, closely related to the molecular gas
mass (see e.g.\ \citealt{bolatto13}).

Third, the other widely used relation exploited here connects the RC
and CO emission. The relation between CO luminosity, \lco, and radio
luminosity, \lrad, has been known since early CO observations (e.g.\
\citealt{rickard77,israel84,murgia02}) and probed in the local
Universe for different type of galaxies (e.g.\ disk-like; dwarfs;
ultraluminous IR galaxies, ULIRGs) using unresolved observations
(global-scale observations; e.g.\ \citealt{adler91, leroy05,liu15}) as
well as in resolved regions down to $\sim 100$\,pc (e.g.\
\citealt{murgia05,paladino06,schinnerer13}).

This article is the first of a series in which we explore the
connection between the RC and the IR and the molecular gas mass 
  ($M_{\rm H_2}$, traced by the CO luminosity) in spatially resolved
galaxies and in galaxies at higher redshifts. We show the use of RC as
a relatively precise proxy for the molecular gas mass in the local
Universe and beyond. We also present an empirical plane among the RC
emission, the IR luminosity and the CO luminosity. We base our work on
large samples of galaxies built from wide-area surveys, resulting in
more statistically robust work than was previously possible.

Throughout the text, we assume a $\Lambda$-CDM cosmology with $H_0=70$
\,km\,s$^{-1}$\,Mpc$^{-1}$, $\Omega_{\rm M} = 0.3$ and
$\Omega_{\Lambda} = 0.7$.

\section{Sample selection}
\label{sec: sample selection}

Our sample is constructed from seven surveys taken from the
literature, all of which have measurements of IR, RC and low-$J$ CO
lines for galaxies at redshifts, $z<0.3$: \smallskip \begin{itemize}

\item VALES \citep{villanueva17,hughes17a,hughes17b,cheng18,molina19}
  comprises 91 galaxies at ($0.02<z<0.35$) taken from
  H-ATLAS\footnote{\herschel\ Astrophysical Tera-Hertz Large Area
    Survey} \citep{eales10}, with 65 galaxies detected
  spectroscopically in low-$J$ CO ($J=1$--0 or $J=2$--1)
  using the Atacama Large Millimetre Array (ALMA) or the Atacama
  Pathfinder EXperiment (APEX). In addition,
  VALES has rich multi-wavelength coverage from UV to far-IR wavelengths
  compiled by the Galaxy And Mass Assembly survey (GAMA;
  \citealt{driver16}), allowing us to measure \lir\ by spectral
  energy distribution (SED) fitting. In order to homogenise the CO
  line measurements to $J=1$--0, we assume a luminosity ratio
  $L^{\prime}_{\rm CO(2-1)} /L^{\prime}_{\rm CO~(1-0)} = 0.85$
  \citep[as tabulated by][]{carilli13}.

\item The xCOLD GASS survey \citep{saintonge17} is a recent upgrade of
  the COLD GASS survey \citep{saintonge11a,saintonge11b} with
  the newest sample called COLD GASS-low. In total, xCOLD GASS
  contains  532 galaxies, both star-forming and passive ellipticals with stellar
  mass in the range $9.0 < \log (M_*/{\rm L}_{\odot}) < 11.5$, in the redshift
  range, $0.01 < z < 0.05$, which have been observed with the IRAM 30-m
  telescope in CO($J=1$--0) and with APEX in CO($J=2$--1). We
    use only the CO ($J=1$--0) data from this survey. The IR data
  comes from the {\it Infrared Astronomical Satellite} ({\it IRAS};
  \citealt{neugebauer84}).

\item \citet{liu15} (hereafter, Liu15) contains 181 local galaxies
  with IR luminosities between $7.8< \log($\lir$/{\rm L}_{\odot})<12.3$, where
  115 are normal spiral galaxies and the rest are ULIRGs. This
    sample contains CO ($J=1$--0) flux measurements from
    high-resolution CO maps published in the literature
    \citep{chung09,young08,kuno07,gao04,helfer03,sofue03} and IR data
  from {\it IRAS}.

\item The APEX Low-redshift Legacy Survey for MOlecular Gas (ALLSMOG;
  \citealt{cicone17}) comprises 97 Sloan Digital Sky Survey (SDSS DR7)
  galaxies at $0.01<z<0.03$, with stellar masses in the range
  $8.5 < \log($\mstar$/{\rm L}_{\odot})< 10.0$, classified as star forming
  according to the BPT diagram and with a gas-phase metallicity 12 +
  $\log(O/H) \ge 8.5$. In addition, this sample contains measurements
  of CO($J=1$--0) from the IRAM 30-m telescope and of CO($J=2$--1) from APEX.
  We use only their CO($J=1$--0) measurements. We
  derive \lir\ from the {\it Wide-field Infrared Survey Explorer (WISE)}
  photometry (from the ALLWISE cataloge, \citealt{wright10}) at
  12\,\mum\ following \citet{cluver14} .

\item The ATLAS$^{\rm 3D}$ survey \citep{cappellari11} contains 260
  galaxies --- a volume-limited sample ($D < 42$\,Mpc) of
  early-type galaxies (ellipticals, E, and lenticulars, S0)
  brighter than $M_K < -21.5$ mag ($\log $\mstar$ \ge  9.78$\,\msun),
  where 56 galaxies have both CO($J=1$--0) and ($J=2$--1) measurements
  from IRAM 30m \citep{young11}, from which we only use the values
    for CO($J=1$--0). For this sample, we derive \lir\
  from {\it WISE} photometry (from the ALLWISE cataloge) at
  12\,\mum\
  following \citet{cluver14}.

\item \citet{leroy05}(hereafter, Leroy05) contains 121 nearby
  ($v_{\rm LSR}\leq 1,000$\,km\,s$^{-1}$) dwarf galaxies with typical
  dynamical  masses $M_{\rm dyn}\leq 10^{10}$\,M$_{\odot}$, optical diameters
  $d_{25} < 5^{\prime}$, and with atomic hydrogen line widths,
  $W_{20}\leq 200$\,km\,s$^{-1}$, selected from the {\it IRAS} Faint Source
  Catalog (FSC) at 60 and/or 100\,\mum. This sample contains 28
  galaxies detected (at S/N $\geq 5$) in CO ($J =1$--0) and another 16
  galaxies marginally detected (S/N $\sim 3$) from the Arizona Radio Observatory's
  12-m telescope. We derive \lir\ for this sample from {\it IRAS} 60-
  and 100-\mum\ photometry, following \citet{orellana17}.

\item \citet{papadopolous12} (hereafter, Pap12) contains 70 local
  ($z<0.1$) IR galaxies ($10<\log($\lir$/{\rm L}_{\odot})<12$) selected
  from the {\it IRAS}
  Revised Bright Galaxy Survey (RBGS; \citealt{sanders03}) --- a
  flux-limited sample (flux density at 60\,\mum $> 5.24$\,Jy).
  Additionally, this
  sample has coverage of CO($J=1$--0) from the IRAM 30-m telescope.
\end{itemize}

\def \sb{S~band}
\def \cb{C~band}
\def \xb{X~band}
\def \kub{Ku~band}

\
\begin{table}
\centering
\setlength{\tabcolsep}{4pt}
\begin{tabular}{ccccccc}
\hline
\hline
GAMA   &\sb           &\cb            & \xb            & \kub          &$\rm\alpha_{radio}$  \\[2pt]
 ID      & [$\mu$Jy]  & [$\mu$Jy]  & [$\mu$Jy]  & [$\mu$Jy]  &          \\
\hline                                                                                                                                                                 
324931 & 510$\pm$57   & ...        & ...        & $138\pm30$ & $-0.81\pm0.10$  \\
543473 & ...          & $291\pm22$ & $170\pm11$ & $154\pm20$ & $-0.71\pm0.10$   \\
491545 & ...          & $347\pm46$ & $213\pm18$ & $164\pm23$ & $-0.82\pm0.13$  \\
 15049 & ...          & $93\pm9$   & $12\pm14$  & $63\pm17$  & $-0.58\pm0.21$  \\
319660 & ...          & $192\pm28$ & $150\pm16$ & $134\pm25$ & $-0.40\pm0.16$  \\
\hline
 323772 & ...        & $629\pm27$ & ...        & ...        & ... \\
 278874 & ...        & $68\pm11$  & ...        & ...        & ...\\
 346900 & ...        & $174\pm15$ & ...        & ...        & ... \\
600656  & ...        & ...        & ...        & $45\pm17$  & ...\\
 210543 & $235\pm32$ & ...        & ...        & ...        & ...\\
 378002 & $391\pm47$ & ...        & ...        & ...        & ... \\
 216973 & $740\pm38$ & ...        & ...        & ...        & ...\\
\hline
\end{tabular}
\caption{New radio continuum data for VALES galaxies (see \S2.1) in the \sb\ ($\nu = 3$\,GHz), \cb\ ($6$\,GHz), \xb\ ($10$\,GHz) and \kub\ ($15$\,GHz) bands. 
The power-law index, $\alpha_{\rm radio}$ ($S\sim\nu^{\alpha_{\rm
    radio}}$) is obtained from these JVLA measurements and we assume
$\alpha_{\rm radio} = -0.7\pm0.3$ for the galaxies with RC obtained
from the literature (see \S\ref{subsec: radio date} for more
details).}
\label{tab:jvla data - complete}
\end{table}

\subsection{Radio data}
\label{subsec: radio date}

In this work, we present new radio data for the VALES sample obtained
from observations taken with the Karl G.\ Jansky Very Large Array
(JVLA; VLA/13B-376, P.I.: E.~Ibar) at 3, 6, 10 and/or 15\,GHz, with
resolutions between 0$\farcs$6 and 4$\farcs$0 (see Table~\ref{tab:jvla
  data - complete}). Additionally we match the VALES sample with the
publicly available Faint Images of the Radio Sky at Twenty-Centimeters
(FIRST; \citealt{white98}) survey at 1.4\,GHz,to obtain a larger
sample in the radio continuum emission.

We compute the radio spectral index $\alpha$
($S_{\nu} \propto \nu^{\alpha}$) in five galaxies from the VALES
survey with multiple JVLA detections. 
To measure the flux densities in galaxies with
multiple radio band measurements, 
we use the Common Astronomy Software Applications (CASA, release
5.0.0)\footnote{$\rm https://casa.nrao.edu$} (\citealt{mcmullin07})
task {\it imsmooth} to degrade the resolution of the images to a
common worst resolution. Then we measure the integrated flux density
using the task {\it imfit}, which allows us to use the same Gaussian
profile in each image.  

For non-VALES galaxies, the RC data were gathered as follows. In the
case of the Liu15 sample, this catalogue includes its own radio data.
For the xCOLD GASS catalogue, we match this sample with the FIRST
survey, with a matching radius of 4$\arcsec$. For the rest of the
samples (Leroy05, Pap12, ALLSMOG and ATLAS$\rm^{3D}$), the
measurements at 1.4\,GHz are taken from the NRAO VLA Sky Survey (NVSS;
\citealt{condon98}) using a matching radius of 20$\arcsec$. To discard
possible mismatches, we use the SIMBAD
\footnote{$\rm http://simbad.u-strasbg.fr/simbad/$} database \citep{wenger00}, and we
corroborate that several of these galaxies have optical minor axis
greater than 60$\arcsec$ and/or that the optical images show no other
dominant galaxies within a radius of 60\,arcsec.

The continuum radio luminosity was obtained in three different ways:
(1) for VALES galaxies with multiple JVLA detections, we $k$-correct
the flux density, $S_{\rm 1.4GHz}$, using the measured radio spectral
index, including error propagation. These spectral indices and errors
were obtained from the best fit based on all the JVLA measurements
available; (2) for galaxies with only one JVLA, FIRST or NVSS
measurement, we estimate the $k$-corrected $S_{\rm 1.4GHz}$
measurement assuming the typical radio spectral index of $-0.7\pm0.3$
(\citealt{ibar09}) expected for star-forming galaxies. In this case,
the error propagation includes the observed errors and the scatter of
the assumed $\alpha$; (3) for galaxies in Liu15 and Leroy05, we use
those directly provided by the surveys. Finally, we calculate the
radio monochromatic luminosity from
$\displaystyle L_{\rm 1.4GHz} = 4\pi D_{\rm L}^2 S_{\rm 1.4GHz}/(1+z),$ where
$D_{\rm L}$ is the luminosity distance and $S_{\rm 1.4GHz}$ is the
$k$-corrected flux density.

To discard any possible contamination from radio-loud active galactic
nuclei (AGN), we matched the entire sample with the 13$\rm^{th}$
Edition of the Catalog of Quasar and AGNs \citep{veron10}, finding
only four galaxies of our final sample to be radio loud, a typical
signature of an AGN. We also matched our final sample with the ALLWISE
survey \footnote{https://irsa.ipac.caltech.edu/cgi-bin/Gator/nph-dd},
where 90 per cent of our sample have measurements at 3.4, 4.5 and
12\,\mum, allowing us to apply the colour-colour criteria defined by
\citet{jarrett11} to identify AGNs. We found five galaxies to be AGNs.
In summary, $\sim 3$ per cent of our final sample have possible
AGN-related contamination of their RC emission. AGN-contaminated
galaxies will often show an excess of 1.4\,GHz emission with respect
to the sample selected as star-forming galaxies. Fig.~\ref{fig:plane}
shows no statistically significance from AGN contamination, confirming
that making corrections for any excess in the RC emission is
unnecessary.

\subsection{Final sample}
\label{subsec: final sample}

\begin{table}
\centering
\setlength{\tabcolsep}{4pt}
\begin{tabular}{cll}
\hline
\hline
Final Sample &  Original    &  Reference        \\        
(\# of galaxies) &  Sample     &           \\ 
\hline 
 30 & VALES       &  \citet{villanueva17}\\
 12 & VALES (APEX)&  \citet{cheng18}     \\ 
 75 & xCOLD GASS  &   \citet{saintonge17}\\
 67 & Liu15       &  \citet{liu15}  \\    
 36 & Pap12       &   \citet{papadopolous12}\\
 28 & ATLAS$^{\rm 3D}$  & \citet{cappellari11}  \\
 17 & ALLSMOG     &  \citet{cicone17} \\
 13 & Leroy05     &   \citet{leroy05}\\
 \hline
\hline
RC &  Original    &  Reference        \\        
(\# of galaxies) &  Sample     &           \\ 
\hline
 12   & VALES   &  This work        \\        
 189  & NVSS    &  \citet{condon98}  \\        
 77   & FIRST   &  \citet{white98}    \\
\hline
\hline
IR &  Original    &  Reference        \\        
(\# of galaxies) &  Sample     &           \\ 
\hline
42         & H-ATLAS &  \citet{eales10} \\        
142        & {\it IRAS}    &  \citet{neugebauer84} \\        
36         & {\it IRAS} RBGS &  \citet{sanders03}   \\     
13         & {\it IRAS} FSC  & \citet{moshir90}         \\        
45         & ALLWISE & \citet{wright10}    \\
\hline
\hline
CO($J=1$--0) &  Original    &  Reference        \\        
(\# of galaxies) &  Sample     &           \\   
\hline
 30 & VALES      & \citet{villanueva17} \\        
 75 & xCOLD GASS & \citet{saintonge17}    \\        
 36 & Pap12      &\citet{papadopolous12} \\  
 28 & ATLAS$^{\rm 3D}$  & \citet{cappellari11} \\        
 17 & ALLSMOG     & \citet{cicone17} \\        
 13 & Leroy05     & \citet{leroy05}  \\        
 67 & Literature  &  \citet{chung09}  \\
     &             &   \citet{young08} \\
     &             &   \citet{kuno07} \\
     &             &   \citet{gao04} \\
     &             &   \citet{helfer03} \\
     &             &   \citet{sofue03} \\
12$^*$   & VALES (APEX)  & \citet{cheng18}         \\        
\hline
\hline
\end{tabular}
\caption{Summary of the surveys and references used to produce our final sample, indicating the number of galaxies having RC, IR and CO data in each case.\newline
$^*$CO data for these galaxies have been converted from CO($J=2$--1)}.
\label{tab:details}
\label{tab:final info}
\end{table} 

Our final catalogue contains 278 galaxies, where 42 are part of the
VALES sample, 67 from Liu15, 75 from xCOLD GASS, 36 from Pap12, 28
from ATLAS$\rm^{3D}$, 17 from ALLSMOG and 13 from Leroy05 (see
  Table~\ref{tab:details} for more details). This sample is
characterised by: (1) CO line measurements with S/N $>5$ for
luminosities ranging from
$\displaystyle 5.78 < \log\left(L^{\prime}_{\rm CO}/{\rm [K\
    km\,s^{-1}\,pc^2]}\right)< 10.35$, (2) radio detections at S/N
$>5$ resulting in a radio luminosity range of
$18.64 < \log\left(L_{\rm 1.4GHz}/{\rm [W/Hz]}\right)< 23.78$, (3)
redshifts, $z\leq 0.271$, with a median value, $z = 0.017$ and (4)
total \lir\ ranging from
$7.3 < \log\left(L_{\rm IR}/{\rm [L_{\odot}]}\right) < 12.14$. While
the total number of galaxies in these seven samples is $> 1,000$, we
discarded the majority of them either because they are not consistent
with our S/N criteria or because they do not have CO, RC or total IR
measurements. Additionally, to avoid aperture inconsistencies amongst
measurements from different studies, we remove from the sample all
galaxies with CO apertures smaller than the Petrosian aperture\footnote{ 
The Petrosian ratio $\rm R_p$ at a radius {\it r} from the center of an object, is the ratio of the local surface brightness in an annulus at {\it r} to the mean surface brightness within {\it r}.  
Then Petrosian radius $\rm r_p$ (Petrosian aperture = $\rm 2r_p$) is defined as the radius as the radius at which the Petrocian ratio equals some limit value, set to 0.2 in our case.} 
in the $r$ band or the optical angular diameter obtained from
NED\footnote{https://ned.ipac.caltech.edu/ }

\section{Results}
\label{sec:results}

\begin{table}
\centering
\setlength{\tabcolsep}{4pt}
\begin{tabular}{l|cccc}
\hline
\hline
Relation    &  Slope        & Intercept         &           &  Pearson   \\[2pt]
            &  m            & b                 & $\sigma$&  coefficient \\        
\hline 
\lrad--\lir  & $0.96\pm0.02$  &  $11.98\pm0.23$   & 0.33     & 0.934        \\
\lco--\lir   & $0.94\pm0.02$  &  $-0.98\pm0.21$   & 0.29     & 0.946        \\    
\lco--\lrad  & $1.04\pm0.02$  &  $-14.09\pm0.21$  & 0.36     & 0.928        \\
\hline
\end{tabular}
\caption{
Best-fit parameters for each relation shown in Fig.\ref{fig:plane} ($\log(y) = m \times \log(x) + b$).
In all these relations, the null hypothesis is rejected.
The $\sigma$ value is measured from the scatter, shown in the respective relations.}
\label{tab:coeff}
\end{table} 

\subsection{The evolution of the RC--CO relation}
\label{subsec:radio-lco-lir plane}

In order to build a RC--CO relation, we first use a RC--IR correlation
that is dependent on redshift, together with the IR--CO correlation.
For the RC--IR correlation, we use the result obtained by
\citet{magnelli15}: \begin{equation} \log \left(\frac{L_{\rm IR}}{\rm
      [W]}\right) =
  \frac{(2.35\pm0.08)}{\left(1+z\right)^{\left(0.12\pm0.04\right)}} +
  \log \left(\frac{L_{\rm 1.4GHz}}{\rm [W\,Hz^{-1}]}\right) + 12.85,
 \label{eq:magnelli}
\end{equation}
where \lir\ is the total integrated IR luminosity (rest-frame
8--1,000\,\mum), $z$ is the redshift, and \lrad\ is the radio
continuum luminosity at 1.4\,GHz. This relation was obtained from a
sample of $\sim340,000$ galaxies at $z < 2$.

For the IR--CO correlation, we use the results obtained from
\citet{villanueva17} based on the VALES sample. They showed that the
global \lco-\lir\ relation can be represented, thus: \begin{equation}
\log \left(\frac{L_{\rm IR}}{\rm [W]}\right) = (0.95\pm0.04)\times\log
\left(\frac{L^{\prime}_{\rm CO}}{\rm [K\,km\,s^{-1}\,pc^{2}]}\right) + (28.6\pm0.4), 
\label{eq:villanueva}
 \end{equation}
\noindent
where the \lco\ is the CO ($J=1$--0) luminosity.

Finally, combining the RC--IR and the IR--CO relations (equations
\ref{eq:magnelli} and \ref{eq:villanueva}), we obtain an expression
for the RC--CO relation as a function of redshift: \begin{equation}
  \begin{aligned}
 \log &\left(\frac{L^{\prime}_{\rm CO}}{\rm [K\,km\,s^{-1}\,pc^{2}]}\right)  =  (2.47\pm0.13)\times(1+z)^{(-0.12\pm0.04)}\\
  & ~~~~~ +(1.05\pm0.04)\times\log\left(\frac{L_{\rm 1.4GHz}}{\rm [W\,Hz^{-1}]}\right)-(16.5\pm0.8).\\
\label{eq:MV}
\end{aligned} 
\end{equation}
Errors were propagated assuming a Gaussian probability distribution.
For simplicity, we define eq. \ref{eq:MV} as $\xi\left(L_{\rm 1.4GHz},z\right)$.

\begin{figure} \centering \includegraphics[bb=68 140 600
  630,width=0.42\textwidth]{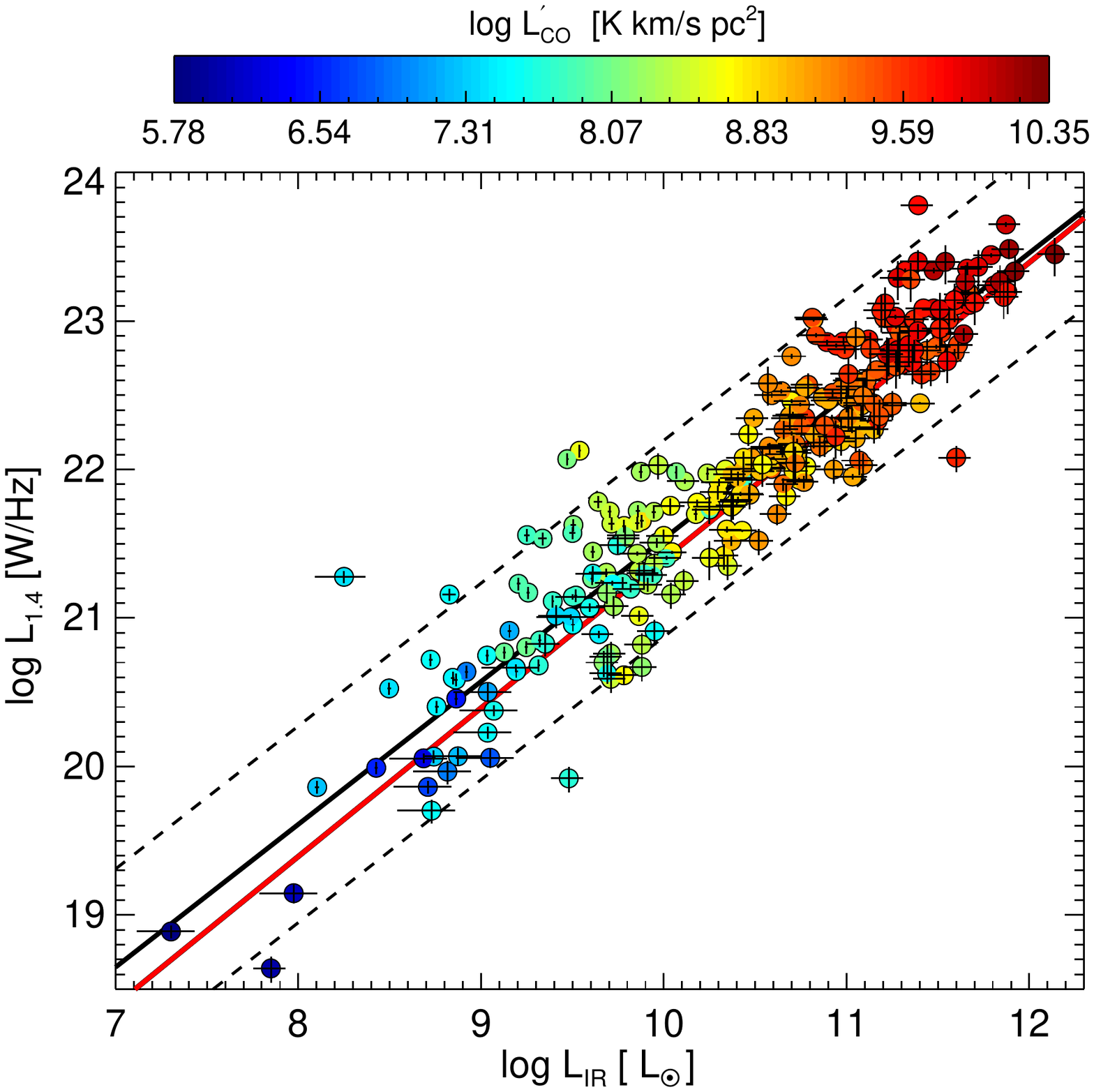}
  \includegraphics[bb=68 140 600
  650,width=0.42\textwidth]{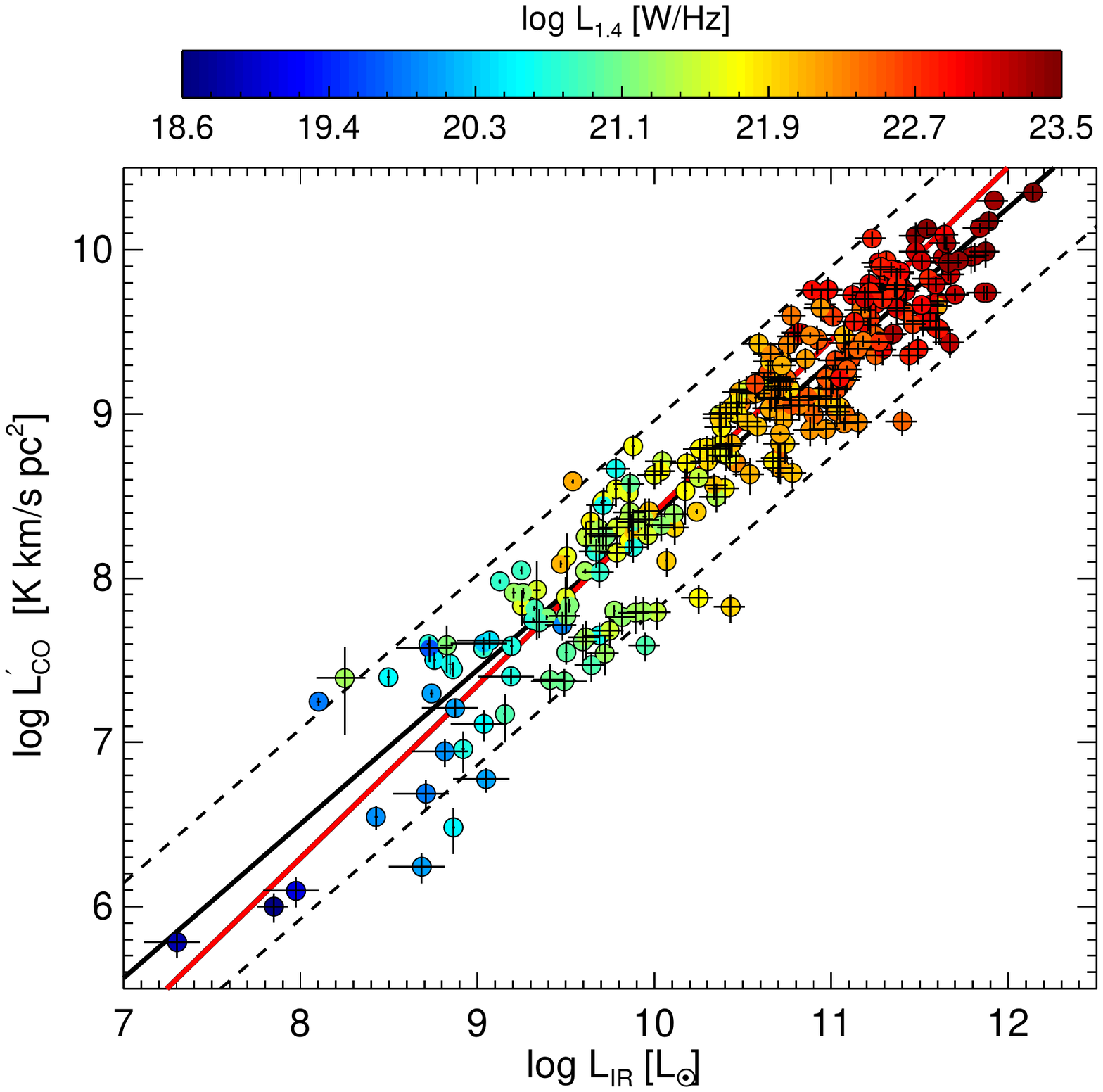}
  \includegraphics[bb=68 170 600
  650,width=0.42\textwidth]{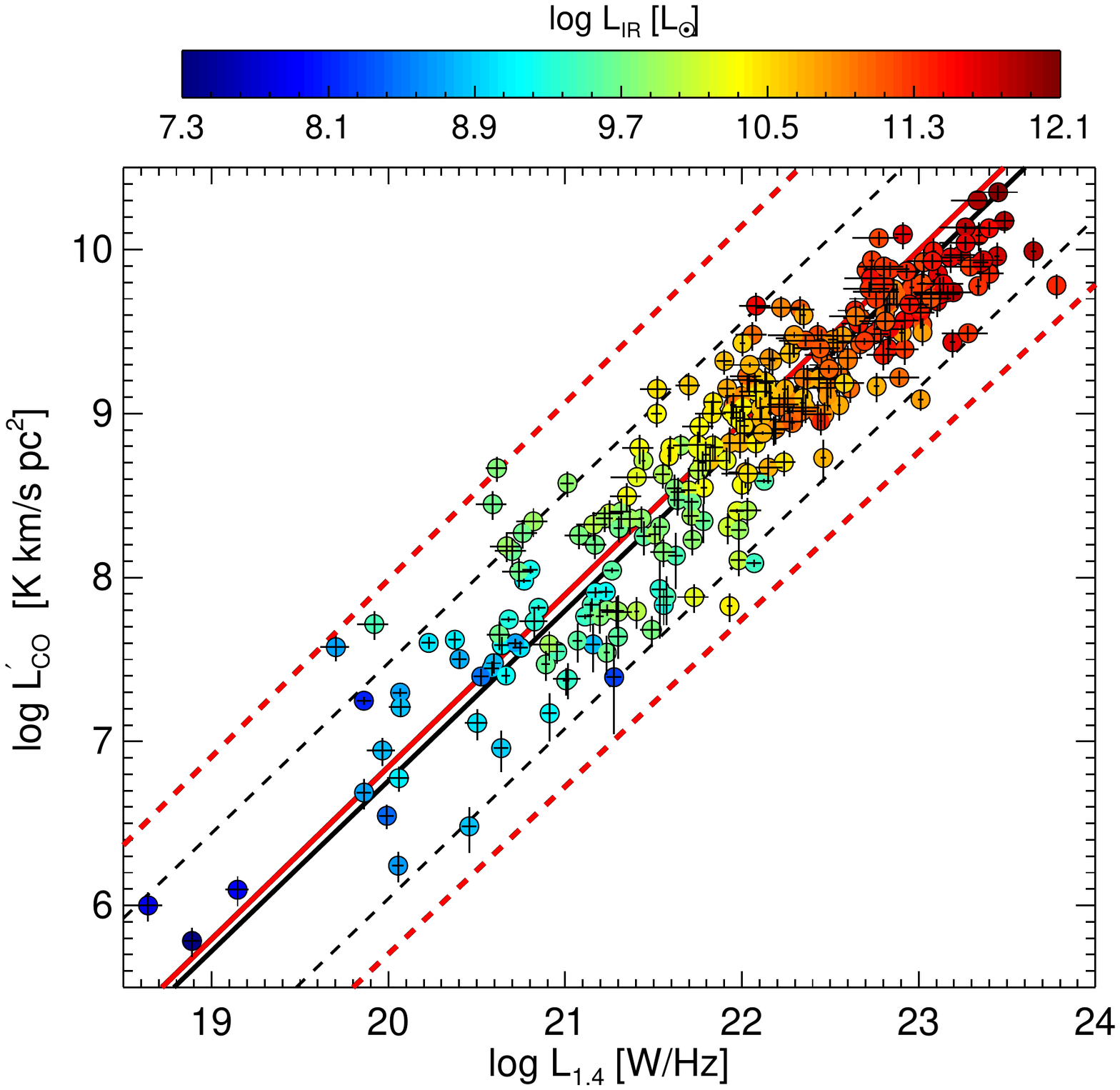}
  \caption{Panels are the \lrad--\lir\ relation with \lco\ in colour
    (top), the \lco--\lir\ relation with \lrad\ in colour (middle),
    and the \lco--\lrad\ relation with \lir\ in colour (bottom). Black
    lines shows the best linear fit (continuous) and 1-$\sigma$
    dispersion (dashed). Red lines are the results from equations
    \ref{eq:magnelli}, \ref{eq:villanueva} and \ref{eq:MV}, for top,
    middle and bottom panels, respectively, assuming the median
    redshift of the sample, $z = 0.017$, for equations
    \ref{eq:magnelli} and \ref{eq:MV}). Red segmented lines in the
    bottom panel shows the error of the relation \lco--\lrad, obtained
    by propagating the errors in 
    in eq.~\ref{eq:MV}.}
  \label{fig:plane}
\end{figure}

Fig.~\ref{fig:plane} shows the RC--IR (top panel), IR--CO (middle
panel) and RC--CO (bottom panel) relations, respectively. Although the
samples are different, our fitted solutions and scatter
($< 0.33$\,dex, see Table~\ref{tab:coeff}) for the RC--IR and the
IR--CO plots are in agreement with the respective results by
eq.~\ref{eq:magnelli} and eq.~\ref{eq:villanueva} (black line).
Additionally, the newly constructed RC--CO relation, based on
equations \ref{eq:magnelli} and \ref{eq:villanueva}, closely matches
the best fit of our sampled data.

\begin{figure*}
\centering
\includegraphics[bb=35 140 580 620,width=0.31\textwidth]{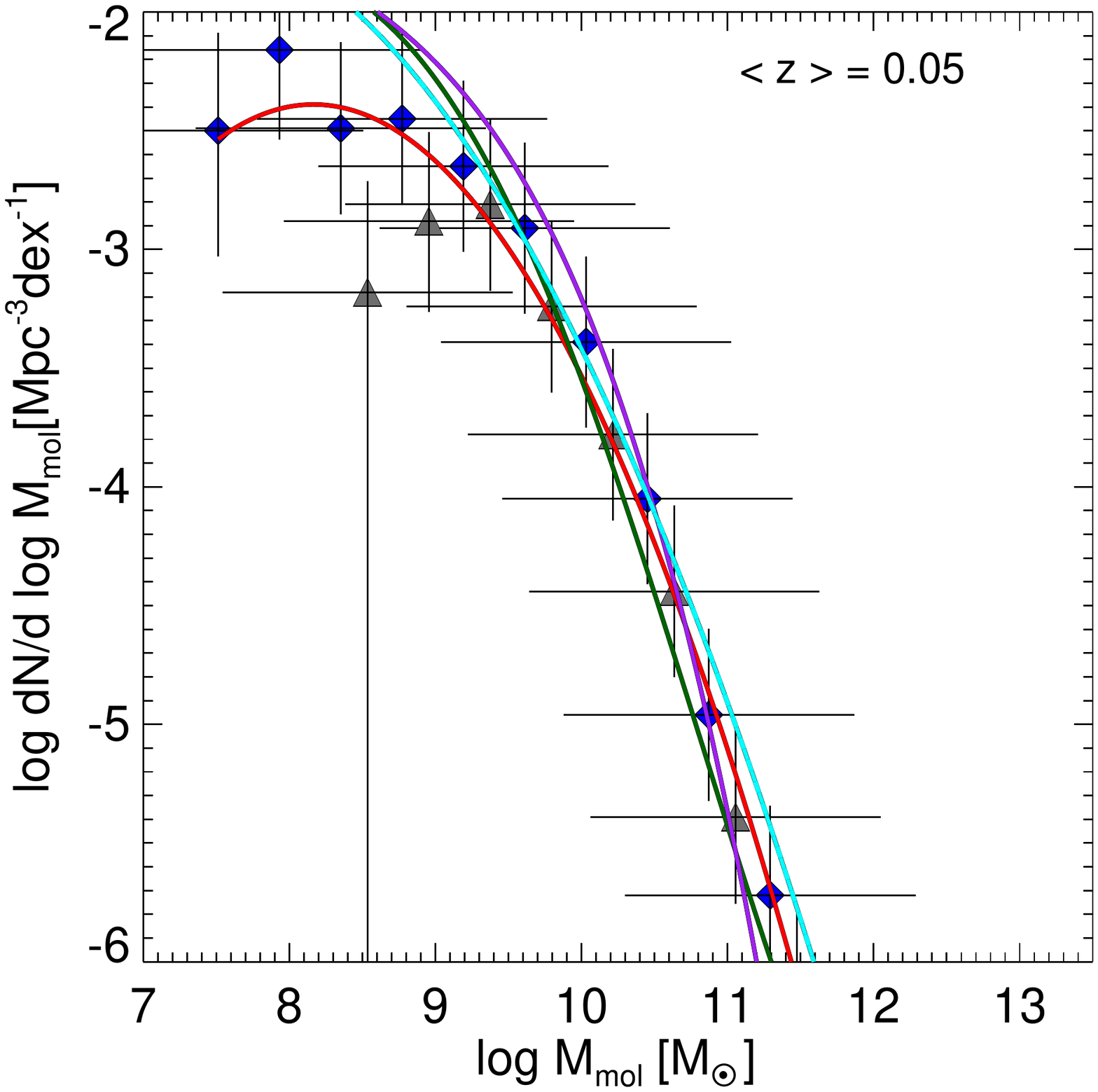}
\includegraphics[bb=35 130 580 690,width=0.31\textwidth]{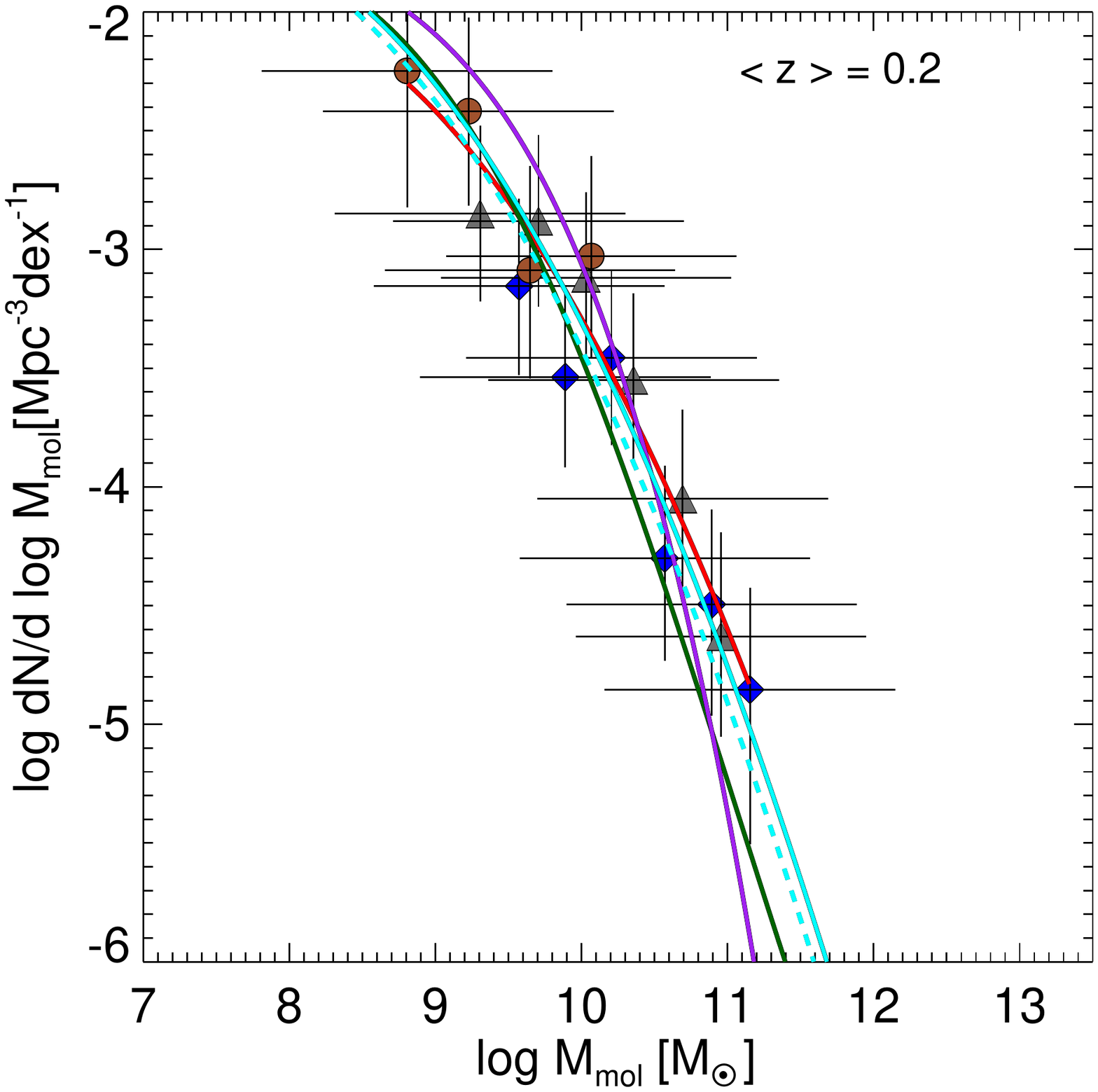}
\includegraphics[bb=35 130 580 690,width=0.31\textwidth]{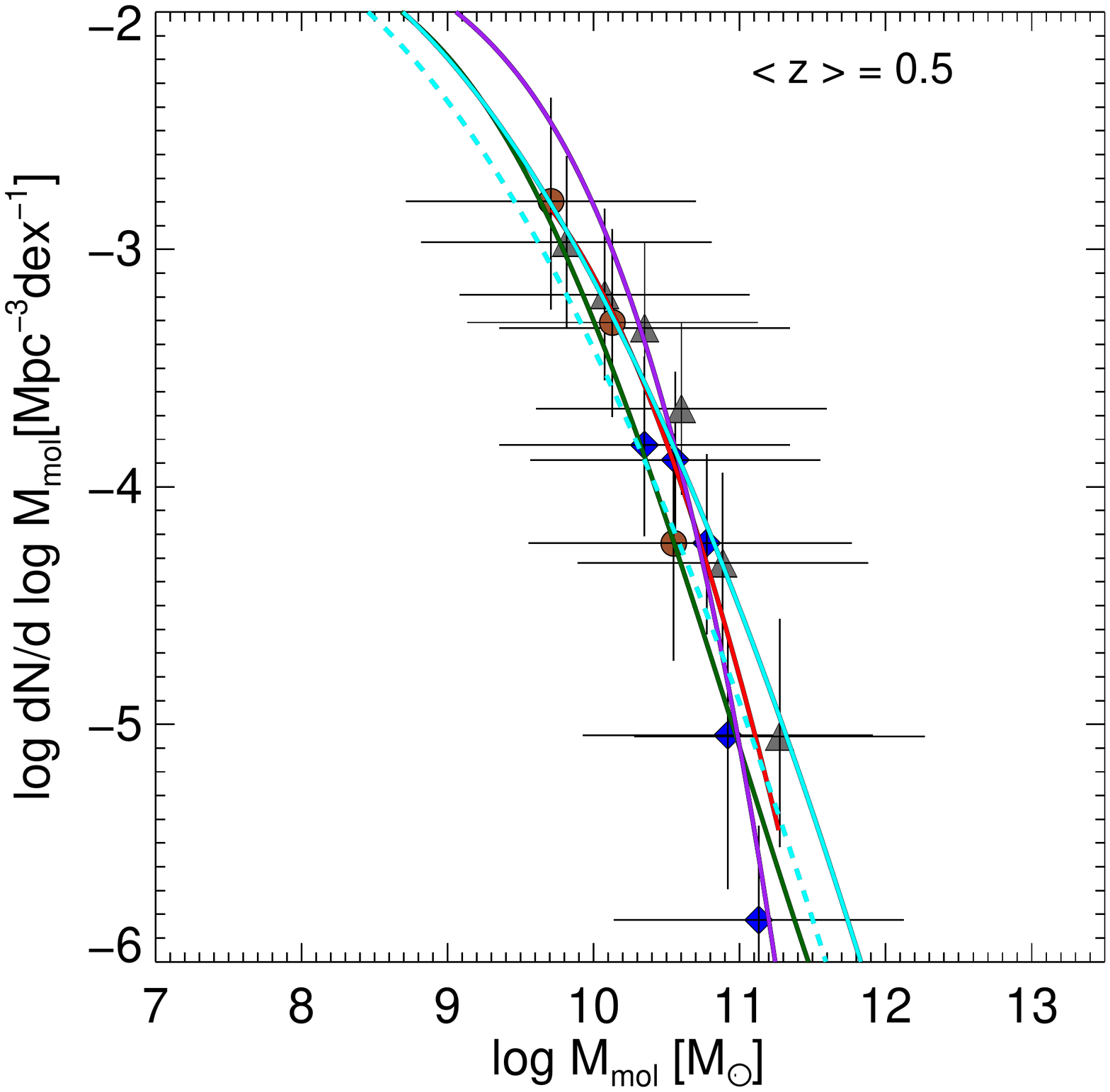}
\includegraphics[bb=35 130 580 690,width=0.31\textwidth]{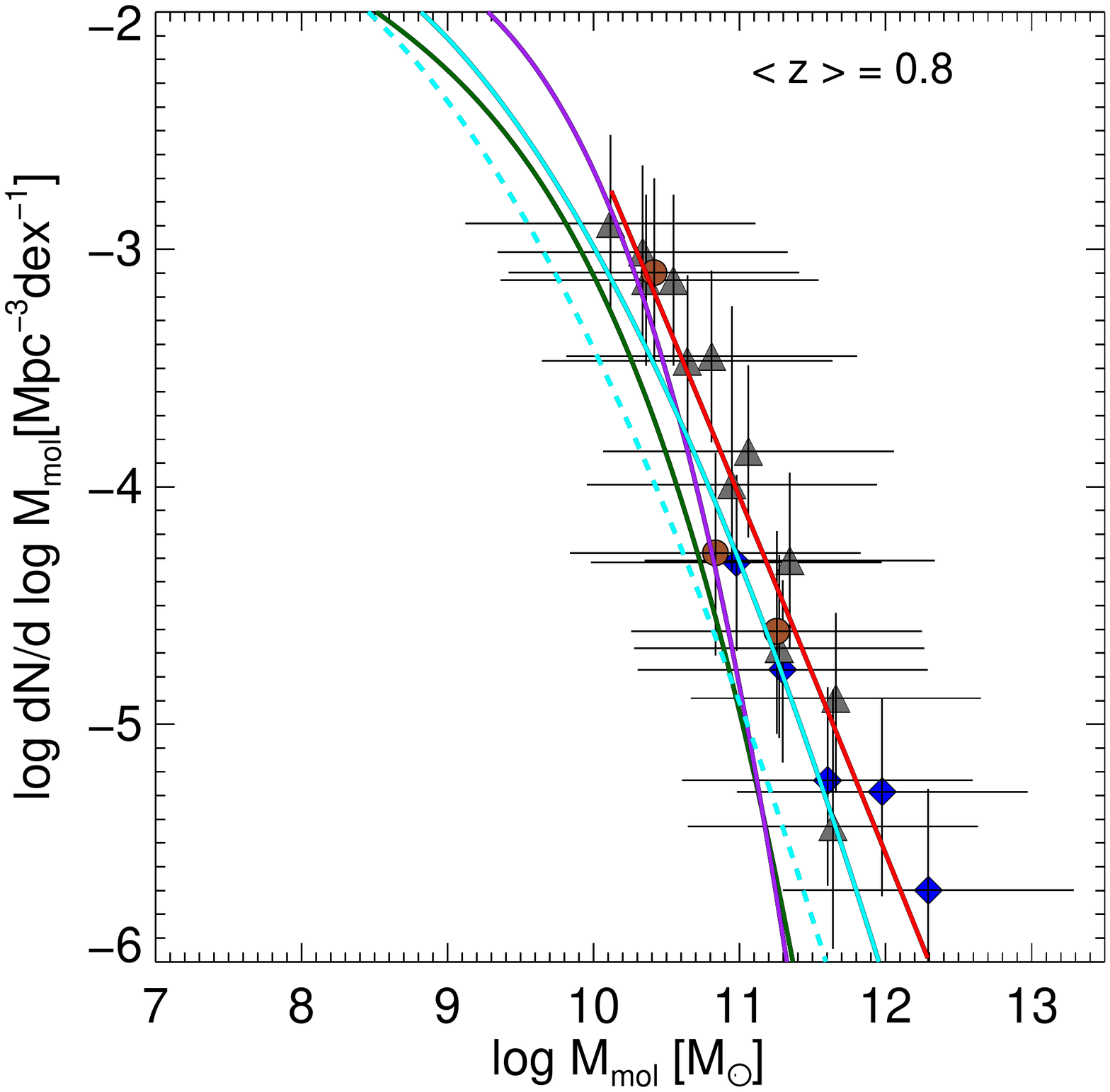}
\includegraphics[bb=35 130 580 690,width=0.31\textwidth]{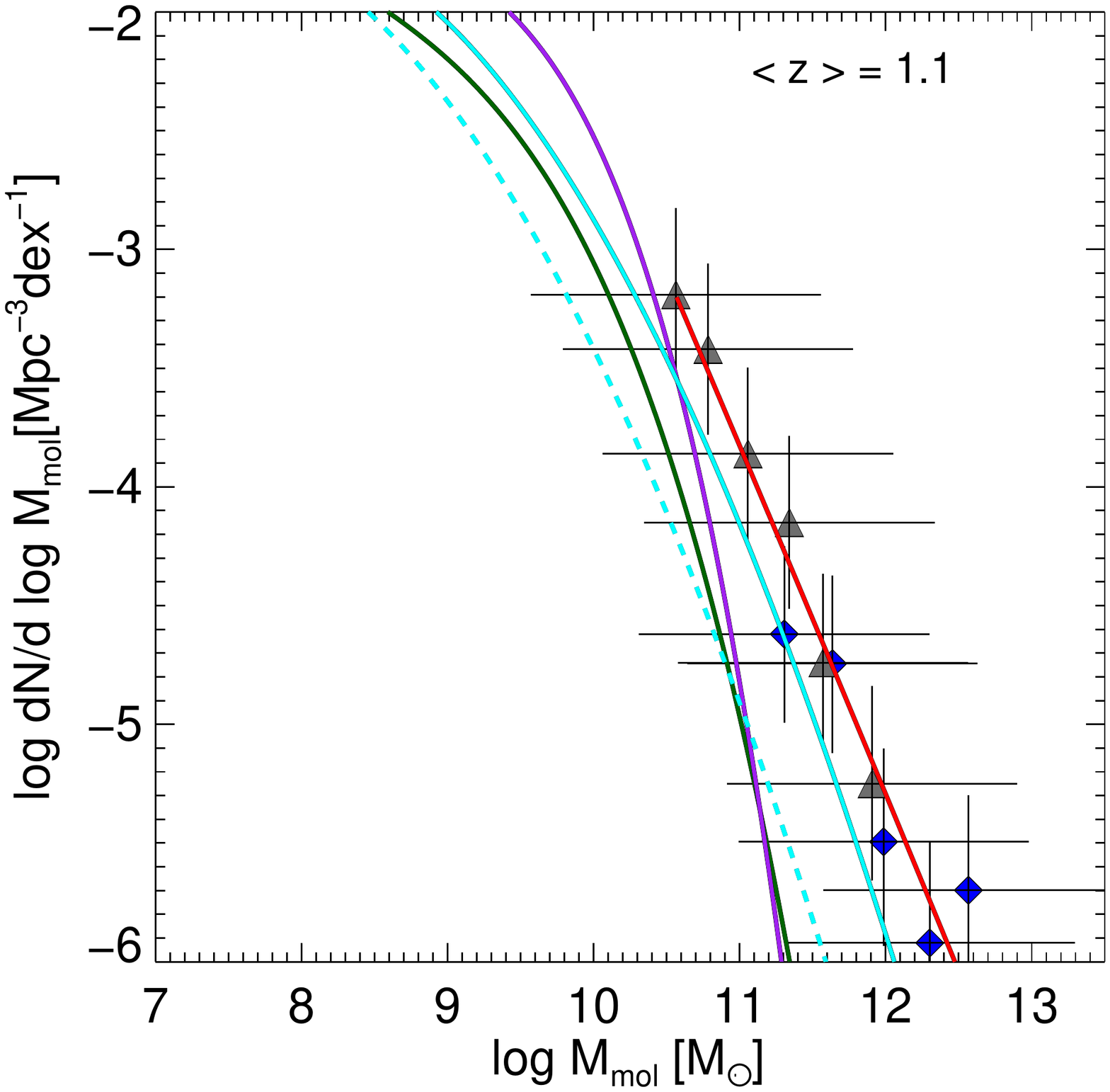}
\includegraphics[bb=35 130 580 690,width=0.31\textwidth]{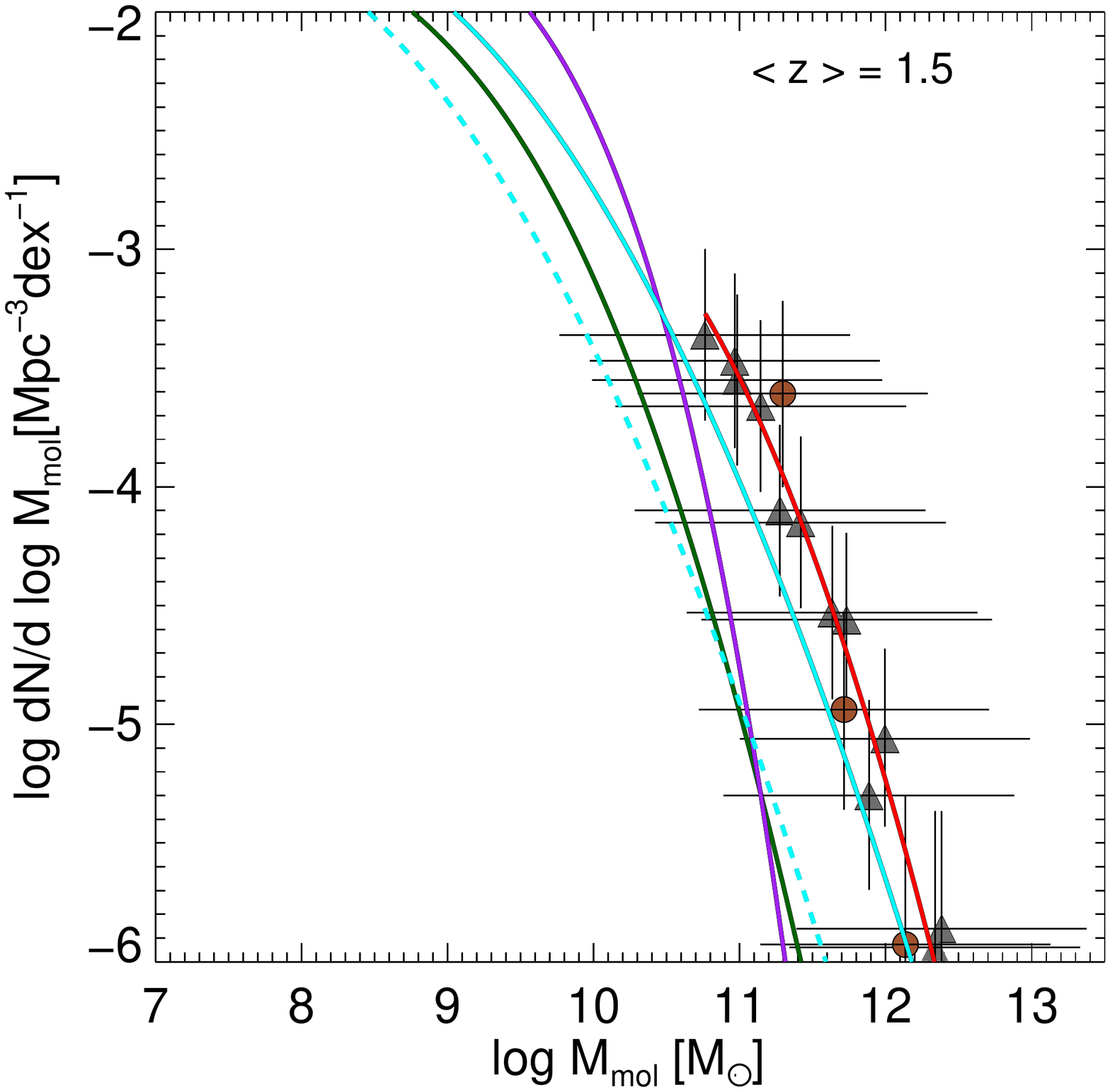}
\caption{Evolution of the MGMF. The upper-left panel shows the local
  ($<z> =0.05$) MGMF for star-forming galaxies, from RCLF measured by
  MS07 (blue diamonds) and Pr16 (gray triangles). The rest of the
  panels show the MGMF for star-forming galaxies at
  $<z> = 0.2, 0.5, 0.8, 1.1$ and $1.5$ from RCLF measured by N017 (red
  diamonds), Sm09 (gray triangles) and Pa11 (brown circles). Red lines
  are the best-fit mass function for all available points, purple and
  dark green lines are the MGMF from the \citet{popping12} and
  \citet{lagos18} models, respectively. Additionally, the cyan solid
  line is the MGMF from the RCLF model from \citet{smolcic09} in
  combination with eq. \ref{eq:MV} and constant
  $\alpha_{\rm CO} = 3.6~{\rm [M_{\odot}/(K\,km\,s^{-1}\,pc^2)]}$ ; similarly, cyan
  dashed lines in all plots correspond to the same MGMF but for a
  constant redshift of $z = 0.05$. The errors in the molecular mass are
  the quadrature addition of the errors from measurements in each
  work and the scatter obtained in our relations. }
\label{fig:molecular lf}	
\end{figure*}

\begin{figure} \centering
  \includegraphics[bb=150 140 430 590,width=0.20\textwidth]{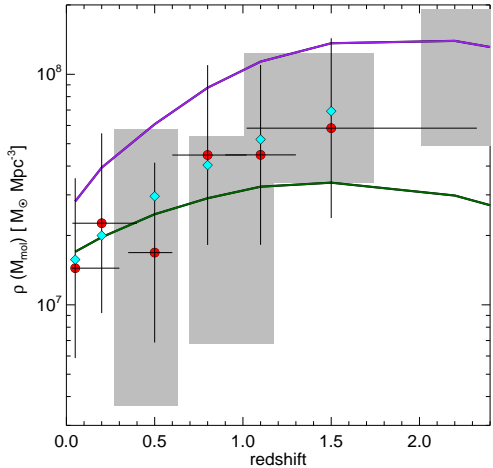}
  \caption{Cosmic
    molecular mass density (\rhom) for the redshift range 0--1.5.
    Red points are the integrated values for the entire mass range
    available for the MGMFs from Fig.~\ref{fig:molecular lf}. Cyan
    points, purple and dark green lines are the integrated MGMF from
    the \citet{smolcic09}, \citet{popping12} and \citet{lagos18} models,
    in the mass range $9 < log(M_{mol}$/\msun)$< 11$,
    respectively. Gray rectangles is the \rhom\ obtained by
    \citet{decarli16}. }
\label{fig:cosmic mol mass}
\end{figure}

\subsection{The RC--CO correlation used to estimate the molecular gas mass function}
\label{subsec: radio coo correlation}

\begin{table}
\centering
\setlength{\tabcolsep}{4pt}
\begin{tabular}{l|cccl}
\hline
\hline
Sample     &  N$^{\circ}$& Redshift   & Alias & References\\[2pt]
           &   bins      & range      &       & \\        
\hline 
NVSS/6dFGS &  1         & 0.003--0.3 & MS07 & \citealt{mauch07} \\
FIRST/SDSS &  1         & 0.005--0.3 & Pr16 & \citealt{pracy16}  \\
VLA/COSMOS &  4         & 0.1--1.3   & Sm09 & \citealt{smolcic09} \\    
VLA/CDF-S  &  4         & 0.03--2.3  & Pa11 & \citealt{padovani11} \\
VLA/COSMOS &  11        & 0.1--5.7   & No17 & \citealt{novak17} \\
\hline
\end{tabular}
\caption{Radio continuum luminosity function surveys.}
\label{tab:LF}
\end{table} 

One of our most relevant results is that we have found a RC--CO
correlation for a large sample as a function of redshift, from
eq.~\ref{eq:MV}. The sample includes galaxies with different Hubble
types at distances greater than 250\,Mpc over a wide range of RC and
CO line measurements. Because the \lrad\ is related to \lco\ which is
in turn related to the molecular gas mass, $M_{\rm mol}$, via a linear
relation, it is possible to estimate the molecular gas mass function
(MGMF, $\phi(M_{\rm mol})$) using RC luminosity function (RCLF,
$\phi(L_{\rm RC})$) measured at different redshifts in samples of
galaxies which exclude the galaxies contaminated by radio-loud AGN.
Then, if we replace $\log L_{\rm 1.4GHz}$ by $\phi(L_{\rm RC})$ in
eq.~\ref{eq:MV}, we obtain the CO luminosity relation as a function of
the redshift. Consequently, by applying eq.~\ref{eq:MV} to
well-measured RCLFs from the literature (see table~\ref{tab:LF}), we
can obtain the CO luminosity functions across the redshift range
$0.003 < z < 5.7$, subdivided into several redshift bins. Although
part of the data taken from literature to construct our final sample
reach almost $z\sim 6$, our results are constrained up to $z=1.5$
based mainly on the redshift constraints included in
eq.~\ref{eq:magnelli} and eq. \ref{eq:villanueva}.

To predict radio luminosity functions
  independently of the availability of radio continuum data, we can
  use the evolution of the RCLF using the model proposed by
\citet{smolcic09}. In this scenario, the local RCLF is based on the
model proposed by \citet{sadler02}, using a combination of a power-law
and a Gaussian distribution of the form: \begin{equation}
\phi_{z=0}\left(L_{\rm RC}\right) = \phi^* \left(\frac{L_{\rm
      RC}}{L_*}\right)^{1-\alpha}
\exp\left[\frac{-1}{2\sigma^2}\left(\log\left[1+\frac{L_{\rm RC}}{L_*}\right]\right)^2\right],
\label{eq:lumfunc} 
\end{equation}
with a power-law index, $\alpha = 0.84$, a luminosity function at the
faint end of $\phi^*=22.9\times10^{-3}\,{\rm [Mpc^{-3}]}$, a standard
deviation of the  Gaussian distribution, $\sigma=0.94$, and a
characteristic luminosity limit between the power-law and the Gaussian
distribution, $L_*=1.95\times10^{19}\,{\rm [W\,Hz^{-1}]}$ (for more details, see \citealt{sadler02,smolcic09}).
Consequently, the evolution of the luminosity function is:
\begin{equation}
    \phi\left(L_{\rm RC}, z\right) = \phi_{z=0}\left(\frac{L_{\rm
          RC}}{(1+z)^{\alpha_{\rm L}}}\right),
    \label{eq:lumfunc evol}
\end{equation}
where $\phi_{z=0}$ is the local luminosity function and a
$\alpha_{\rm L}$ is the characteristic luminosity function parameter,
with value $\alpha_{\rm L}=2.1\pm0.2$ \citep{sadler02}.

\lco\ is calculated using eq. \ref{eq:MV} and is used to obtain the CO
luminosity functions, transforming equations \ref{eq:lumfunc} and
\ref{eq:lumfunc evol} as follows: \begin{equation} \begin{aligned}
 \phi_{z=0}\left(L^{\prime}_{\rm CO}\right)  &=
 \xi\left(\phi_{z=0}\left(L_{\rm 1.4GHz}\right)\right) \\
 \phi\left(L^{\prime}_{\rm CO},z\right) &= \xi\left(\phi\left(L_{\rm 1.4GHz}, z\right)\right)
\end{aligned}
\end{equation}

Finally, to obtain the MGMF, we use the conversion factor between
\lco\ and the molecular gas mass, $M_{\rm mol} =\alpha_{\rm CO}$ \lco,
with constant \lco--$\alpha_{\rm CO}$, where $\alpha_{\rm CO}$ is the
molecular mass conversion factor and has a value
3.6\,M$_{\odot}$/(K\,km\,s$^{-1}$\,pc$^2$). Although this value for
$\alpha_{\rm CO}$ does not take into account the helium contribution
(the fraction of He is 36 per cent), we adopt it considering
that the reference sample to compare our resulting luminosity
functions, \citet{decarli16}, uses this value.

We compare our results with those obtained from the semi-analytic
  MGMF models developed by \citet{popping12} and \citet{lagos18}.
Fig.~\ref{fig:molecular lf} shows the MGMFs (red and cyan lines)
estimated for six redshift bins centred at
$<z> = 0.05, 0.2, 0.5, 0.8, 1.1$ and $1.5$, covering the redshift
range from 0.003--2.3. Each sample used is identified by different
symbols and colours (see the caption to Fig.~\ref{fig:molecular lf}).
The errors result from the addition in quadrature of the error
from the original RCLF and the scatter in the \lco--\lrad\
correlation. The purple \citep{popping12} and dark green
\citep{lagos18} lines corresponds to semi-analytic models.

Although we could potentially construct a RCLF up to $z=5.7$,
our results for the MGMF show a considerable over-estimate in
comparison with the semi-analytic models for RCLF at redshifts
$z > 1.5$.
This problem can be caused by: a) the \lco--\lrad\ correlation
function evolving with redshift in a different way than that
considered in this work; b) it may be incorrect to adopt a constant
$\alpha_{\rm CO}$ given that starburst galaxies are more abundant at
higher redshifts; c) the RCLM at higher redshifts only characterises
the galaxies richest in molecular gas, as a consequence of the
detection limit in the radio images; d) semi-analytic models may
under-estimate the molecular gas mass at high redshift. 
Additionally, with the assumption of a constant $\alpha_{\rm CO}$,
  we ignore any possible variation of the conversion factor as a
  function of galaxy properties, where the most important property is
  the dependence on gas-phase metallicity (e.g.\
  \citealt{wilson95,arimoto96,barone00,israel00,boselli02,
    magrini11,schruba12,hunt15,amorin16}).

\subsection{Cosmic molecular gas mass content}
\label{subsec: codmic moleculat gas mass}

Based on the results obtained from our MGMF analysis, we can predict
the molecular gas mass density and its evolution with cosmic time. Molecular gas mass density is defined as:
\begin{equation}
\rho_{(M_{\rm mol})}=\int^{\log M_2}_{\log M_1} \frac{dN}{d\log M~ dV} ~ M~ d\log M.  
\end{equation}
Fig.~\ref{fig:cosmic mol mass} shows the resulting \rhom\ in the
redshift range $0 < z < 1.5$, where the red points are derived using
the entire mass range available from the respective RCLF. This means
that this derived \rhom\ has different integration mass limits in
different redshift bins. For this reason, we make use of the RCLF from
\citet{smolcic09} to obtain a continuous MGMF where we can extrapolate
a fixed integration mass limit to compare our results with the
predictions from semi-analytic models. After fixing the integration
mass limit in the range $9 < \log(M_{\rm mol}/{\rm M}_\odot)< 11$ (the
range deduced from the MGMFs obtained by \citealt{decarli16}, but not
specified in that work) we can then determine the \rhom\ based on our
continuous MGMF (cyan diamonds). Our results fall inside the errors of
the direct measurements obtained by \citet{decarli16} (gray
rectangles), while the semi-analytic models show higher and lower
values for the models from (\citet{popping12} (purple line) and
\citet{lagos18} (green line), respectively.

The main source of uncertainty in \rhom\ comes from the uncertainty in
the measured RC luminosity functions.

\subsection{The empirical CO--RC--IR plane}
\label{subsec: final plane}

\begin{figure} \centering
  \includegraphics[width=0.45\textwidth]{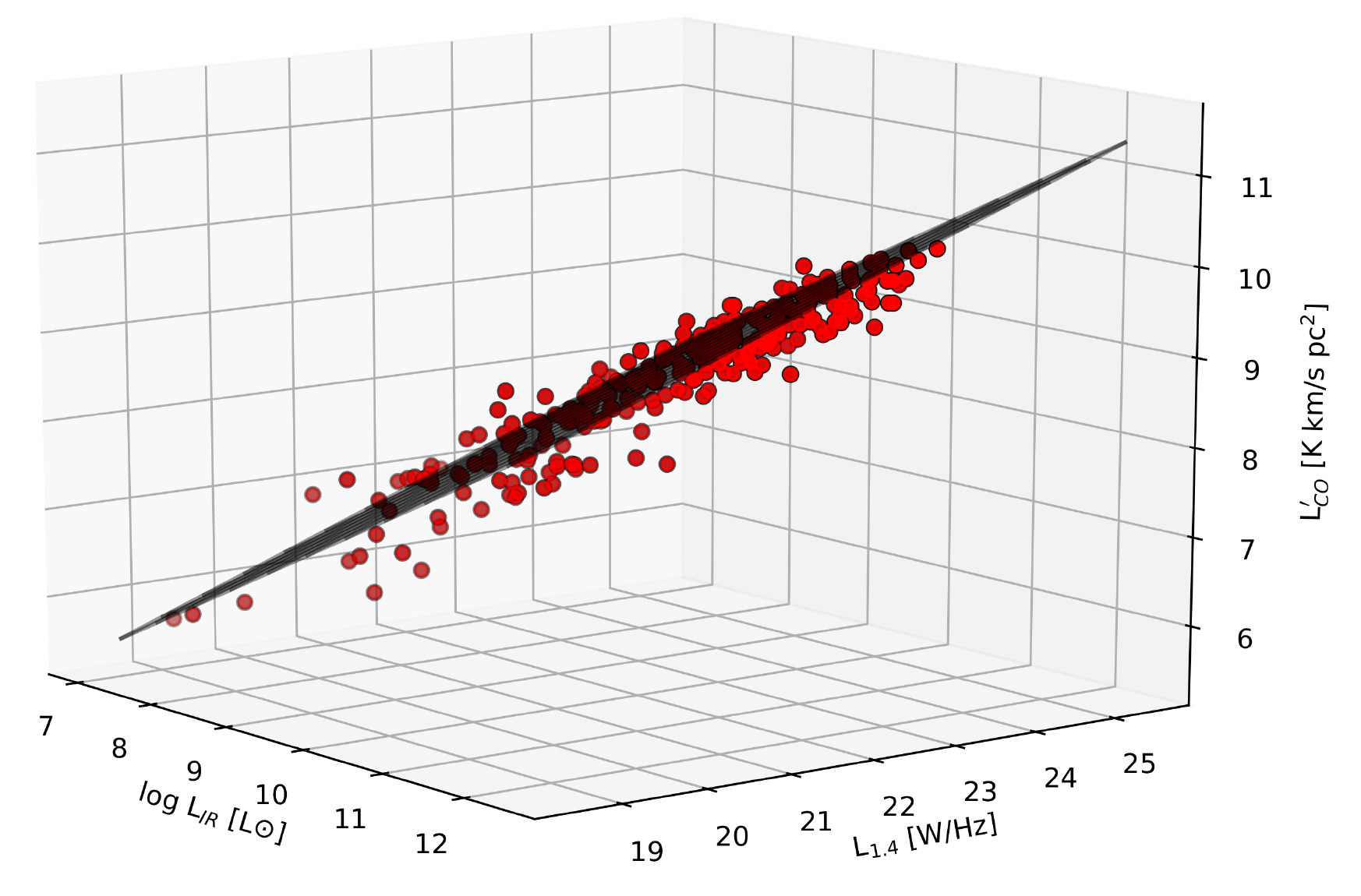}
  \caption{The CO--RC--IR plane. Red circles are the galaxies from
    the final sample and the black plane is the best fit shown in
    eq.~\ref{eq:plane}. See more views in Appendix~\ref{ap:A}}
  \label{fig:plane-final_one} \end{figure}

Although the physical parameters used for the star-formation relations
in this paper --- the IR, CO and RC emission --- are all tracers of
newly formed stars, to our knowledge there is no previous use in the
literature of parameter space combining all of them in a single
relation. Fig.~\ref{fig:plane} shows that these three parameters vary
in a correlated way. Based on that, we conclude that each panel
corresponds to the projection of a CO--RC--IR plane. We present the 3D
view of these three relations in Fig.~\ref{fig:plane-final_one}
forming a plane for which we obtained a numerical expression using the
MPFIT
algorithm\footnote{https://www.physics.wisc.edu/$\sim$craigm/idl/fitting.html}
\citep{markwardt09}:

\begin{equation}
\begin{aligned}
 \log&\rm\left(\frac{L^{\prime}_{\rm CO}}{\rm
     [K\,km\,s^{-1}\,pc^2]}\right) = (0.61\pm0.05)
 \times\log\left(\frac{L_{\rm IR}}{\rm [L_{\odot}]}\right) \\
 & ~~~~~~~~+ (0.36\pm0.05) \times\log\left(\frac{L_{\rm 1.4GHz}}{\rm [W\,Hz^{-1}]}\right) - (5.4\pm0.7) ~. 
\label{eq:plane}
\end{aligned}
\end{equation}
\noindent

For different views of the plane based on eq.~\ref{eq:plane}, see Appendix~\ref{ap:A}.

\section{Discussion}
\label{sec:conclusion}

\subsection{Cosmic evolution of molecular gas}

The intimate connection between the RC, CO and IR emission during the
process of forming stars is well established for samples of galaxies.
However, understanding the mechanism that links these three parameters
remains a challenge for star-formation models, considering the
different physical origins for the different emission.

Several tracers of star-formation activity exist. Direct methods
  include tracing the UV light from young stellar populations.
  However, UV attenuation by dust in galaxies requires additional
  corrections. For example, it
  has been proposed that the emission at 24\,\mum\ can be used to
  correct the dust attenuation in far-UV (FUV) and H$\alpha$ data,
  resulting in more realistic estimates of the SFR in galaxies
  \citep{kennicutt07,calzetti07}. These combined calibrations of SFR
  are used extensively for nearby galaxies (e.g.\
  \citealt{rahman11,ford13,momose13,casasola17})

To avoid correcting for extinction, IR emission is widely used as
  a star-formation tracer. Two energy sources contribute to the IR
spectral energy distribution (SED): a warm dust component, where
grains absorb UV photons from young, massive ($> 8$\,\msun) stars
(e.g.\ \citealt{devereux91,condon92}), and a cold dust component --- a
consequence of the absorption of optical photons from the interstellar
radiation field (e.g.\ \citealt{xu96}). The IR emission is a powerful
star-formation tracer, better than other more direct indicators that
are affected by extinction (e.g.\ H$_\alpha$\ emission systematically
under-estimates, by an order of magnitude, the formation of stars in
galaxies with SFR $\geq 20$\,\msun\,yr$^{-1}$ \citealt{cram98}).

The RC emission results from the combined contribution of thermal
emission --- free-free radiation from H\,{\sc ii} regions related to
the number of massive, short-lived stars --- and non-thermal emission
--- synchrotron radiation produced by relativistic electrons
interacting with the magnetic field of the galaxy \citep{condon92}.
These two components dominate the RC at different frequencies, where
the transition between the thermal and non-thermal component happens
at $\nu \approx 30$\,GHz (emission at higher frequencies is mainly
thermal and at lower frequencies is mainly synchrotron).
\citet{condon92} suggests that thermal radiation contributes less than
10 per cent of the total RC radiation at $\sim 1$\,GHz and that the
total radio luminosity at 1.4\,GHz is directly proportional to the
supernovae (SN) rate. Synchrotron radio emission from star-forming
galaxies is produced by the interaction of energetic electrons ---
accelerated by the massive stars when they have reached the supernovae
stage --- and the ambient magnetic field \citep{israel84,voelk89}.
Radio emission is unaffected by dust absorption, making RC emission an
ideal tool to calibrate other tracers of star formation.

Lastly, CO emission is the consequence of the interaction between CO
molecules and the molecular hydrogen (H$_2$) in regions where stars
are formed --- giant molecular clouds (GMC). CO emission is correlated
with the virial mass of GMCs observed in the Milky Way and in nearby
spiral galaxies \citep{young91}.

Historically, the IR--RC--CO parameters have been studied using three
empirical relations, each one having different properties: first, the
RC--IR correlation (e.g.
\citealt{yun01,ivison10,thomson14,magnelli15,dumas11,tabatabei13a,tabatabei13b}),
valid up to $z \sim 2$ and at a resolution of $\ge 500$\,pc. Second,
the Schmidt-Kennicutt (SK) IR--CO law \citep{kennicutt98,kennicutt12},
valid for high-SFR, bright galaxies as well as low-surface-brightness
galaxies with low SFR. The form of the SK law is still debated because
the observations of disk and starburst galaxies have shown that this
relation can be fitted either by a single or a bi-modal solution for
the redshifts range 0.05--2.5 (e.g.\
\citealt{genzel10,daddi10,ivison11,kennicutt12,tacconi13,santini14,freundlich19}).
The slope of the SK law, which defines the depletion time of the gas
in a galaxy ($\tau_{\rm dep} \equiv M_{\rm gas}/{\rm SFR}$), is
central to our understanding of the mechanisms that govern star
formation. According to models, a linear slope implies that
star-formation activity is not driven only by the self gravity of the
galaxy \citep{semenov17,semenov19}. Several studies dedicated to
  the spatially-resolved KS relation on sub-kpc scales found a wide
  range
  in the value of the slope (indicated by the power law index, $N \sim
  0.6$--3; $\Sigma_{\rm SFR}\propto\Sigma_{M(\rm H_2)}^N$) of the KS
  relation \citep{bigiel08,kennicutt12,rahman11,viaene14,casasola15}.
  The spread in the value of $N$ may be intrinsic, suggesting that
  different SF relations exist; alternatively, it may be due to the
  assumptions adopted in each study.  \citet{daddi10} suggest that
\lir\ (tracing SFR) and \lco\ (tracing molecular gas mass) show no
evolution in normal star-forming galaxies up to $z=1.5$. However,
\citet{tacconi18} suggest that $\tau_{\rm dep} \propto (1+z)^{-0.62}$,
implying that $\tau_{\rm dep}$ decreases by a factor of
$\approx2\times$, up to the redshift limit explored in our study,
$z= 1.5$. Third, the RC--CO correlation (e.g.\
\citealt{adler91,murgia02,murgia05,leroy05,paladino06,schinnerer13,liu15}),
is valid for different galaxy types and also down to scales of few
hundred parsecs in some selected local galaxies.

In this work we derive eq.~\ref{eq:MV} by combining RC--IR and IR--CO
relations from the literature to obtain the molecular gas mass density
of the Universe, after deriving the integrated MGMF from measurements
of the RCLF corrected by
$\alpha_{\rm CO}=3.6~{\rm [M_{\odot}/(K\,km\,s^{-1}\,pc^2)]}$. We find
that the resulting molecular gas mass density increases by a factor of
$4.4\times$ across $z=0.05$--1.5. As mentioned above, we expect that
the evolution of the depletion time has little impact on our final
gas mass densities.

\subsection{Using the CO--RC--IR plane to predict CO emission in galaxies}

 Several authors have proposed models to explain the
  physical connection between the radio, infrared and CO
  luminosities. One of the earliest models proposes that the RC--IR
correlation is physically linked to the formation of young massive
stellar populations (e.g.\ \citealt{helou85,condon92}) since GMCs
collapse to form stars of large masses, responsible for an intense
radiation field in star-forming galaxies. A large fraction of the UV
photons from the forming stars heat the dust grains, re-processing the
UV radiation into IR emission. Considering that the main mechanism
that drives this model (\citealt{adler91,murgia02}) is the formation
of new stars, it is expected that \lco\ --- a classical tracer of the
gas reservoir for future star formation --- is correlated with IR and
RC emission. As a first approach, this model offers a simple mechanism
to explain how these three parameters can be related by considering
that the origin of the IR emission and the non-thermal synchrotron
corresponds to newly formed stars in molecular clouds. However, this
model is unable to explain several problems, e.g.\ why the RC--IR
correlation is maintained at scales of kpc and the RC--CO relation
holds at scales of $\sim100$\,pc. In addition, \citet{schinnerer13}
argue that this model requires too many intermediate physical
processes to observationally correlate both the RC and the IR
emission, where each of these processes has to contribute to a larger
scatter than that obtained from observations. Alternative models,
e.g.\ the calorimeter model (\citealt{voelk89,lisenfeld96}), the
magnetic field-gas density coupling (\citealt{helou93,niklas97}), and
the proton calorimeter model \citep{suchkov93,lacki10} also struggle
to explain the physical mechanisms involved in these relations.

Our work provides an empirical solution that combines CO--RC--IR into
a plane, which can be exploited to update star-formation models. We
find that our proposed plane --- where the CO, RC, and IR luminosities
are fitted simultaneously --- results in smaller scatter for each of
the relations than when they are fitted independently in pairs. For
example, the comparison between the estimated \lco\ obtained using
eq.~\ref{eq:plane} and the actual measured \lco\ shows a scatter of
0.27\,dex (a factor $1.86\times$). The scatter between our sample and
the IR--CO and RC--CO relations are 0.29 and 0.36\,dex, respectively
(see Table~\ref{tab:coeff}). As a consequence, the CO--RC--IR plane
constitutes a powerful tool to predict, in a very efficient way, the
CO emission in large samples of galaxies lacking molecular mass
measurements. Although the CO--RC--IR plane is fitted using a wide
range of galaxy types (local LIRGs, massive star-forming galaxies,
elliptical galaxies with star formation, dwarf galaxies), the scatter
for the predicted CO emission from the plane is lower than the scatter
from independent use of the IR--CO or RC--CO relations.

Using a sample of $\sim80$ metal-rich and -poor dwarf galaxies,
\citet{filho19} studied the star-forming relations RC--IR, RC--CO and
IR--CO finding that they cannot be extended down to low radio
luminosity dwarfs. A breakdown for the dwarfs appears towards brighter
radio luminosities in both relations with respect to the solution for
bright galaxies from \citealt{price92,liu10,kennicutt11,murgia05}.
According to \citet{filho19}, the breakdown in the RC--CO and IR--CO
relations reflects a depletion of CO in dwarf galaxies, which have a
hard ionising radiation field, low dust shielding and slower chemical
reaction rates, making this molecule an inefficient tracer of
molecular gas. Cosmic rays from starburst episodes also play a role in
destroying the CO molecules. We cannot directly compare
\citet{filho19}'s results with our work because the range of
luminosities for their sample of dwarfs is about an order of magnitude
fainter than ours.

Considering that each parameter used in the construction of the plane
involves observed emission coming directly or indirectly from the
formation of stars, the simplest scenario to explain the
existence of this three-fold  relation lies with models where
  star formation is the main driver. Assuming that star formation is
the main driver behind the observed consistency of RC, IR, CO
emission, then implementing it to predict CO emission on a different
sample requires the selection of star-forming galaxies only, where no
contribution of AGN is present.

\section{Summary}

In summary, we have used 278 nearby ($z < 0.27$) galaxies with
measurements of global CO lines ($J=1$--0 or $J=2$--1), total IR
luminosity and radio continuum flux density at 1.4\,GHz to study the
scaling relations between the CO, RC, and IR luminosities and to
determine the cosmic evolution of the molecular gas content of the
Universe. Assuming that star formation is the main mechanism that
drives the existence of the SK (IR--CO) and RC--IR relations, we
derive the RC--CO relation as a function of redshift. This relation
allows us to estimate the molecular gas mass function (MGMF) using the
radio continuum luminosity function (RCLF) from the literature,
obtaining consistent results as compared with the MGMF from
semi-analytic models across $0.05 <z < 1.5$. Finally, from the
integration of the different MGMFs, we estimate the cosmic evolution
of the molecular gas mass density ($\rho(M_{\rm mol})$) in
star-forming galaxies, in six redshift bins, showing an increment of
$4.4\times$ across this redshift range.

In addition, we found that these three luminosities (\lco, \lrad\ and
\lir) form a plane, presumably as a consequence of the common physical
mechanisms behind these three observable quantities in the context of
star formation in galaxies. The plane is valid across more than five
orders of magnitude for each of the luminosities, across the redshift
range 0.05--0.27, and it can be used to predict CO emission from the
IR and radio continuum measurements of galaxies, with a scatter of
0.27\,dex. The agreement of our estimates with semi-analytical models,
as well as with the results from the ALMA large programme, ASPECS, is
a clear indication of the power of the method presented here as a tool
to predict molecular gas mass at different cosmic epochs. 
  Future large radio surveys, such as those conducted by the
  Square Kilometer Array, will be able to exploit these
  results.

\section*{Acknowledgements}

We thank Claudia Lagos and Gerg$\rm\ddot{o}$ Popping, who
provided us with data from their semi-analytic models. The authors
acknowledge support provided by: FONDECYT through grant
N$^\circ$\,3170942 (G.O.G.) and grant N$^\circ$\,1171710 (E.I.);
CONICYT-PIA ACT N$^\circ$\,172033 and CONICYT QUIMAL N$^\circ$\,160012
(R.L.); the Young Researcher Grant of National Astronomical
Observatories, Chinese Academy of Science and the National Natural
Science Foundation of China, N$^\circ$\,11803044 and
N$^\circ$\,11933003 (C.C.); STFC (ST/P000649/1) (A.T.); and the
Chinese Academy of Sciences (CAS) and the National Commission for
Scientific and Technological Research of Chile (CONICYT) through a
CAS-CONICYT Joint Postdoctoral Fellowship administered by the CAS
South America Center for Astronomy (CASSACA) in Santiago, Chile
(T.H.); CONICYT (Chile) through Programa Nacional de Becas de
Doctorado 2014 folio 21140882 (P.C.C.). We acknowledge the National
Radio Astronomy Observatory, a facility of the National Science
Foundation operated under cooperative agreement by Associated
Universities, Inc. 
 This research has made use of the NASA/IPAC
Extragalactic Database (NED), which is funded by the National
Aeronautics and Space Administration and operated by the California
Institute of Technology. 
This research has made use of the SIMBAD database,
operated at CDS, Strasbourg, France 
This work makes use of JVLA project,
13B-376. ALMA is a partnership of ESO (representing its member
states), NSF (USA) and NINS (Japan), together with NRC (Canada), MOST
and ASIAA (Taiwan), and KASI (Republic of Korea), in cooperation with
the Republic of Chile. The Joint ALMA Observatory is operated by ESO,
AUI/NRAO and NAOJ. This paper makes use of the following ALMA data:
ADS/JAO.ALMA\#2012.1.01080.S and ADS/JAO.ALMA\#2013.1.00530.S. This
publication is based on data acquired with the Atacama Pathfinder
Experiment (APEX): programmes 097.F-9724(A) and 098.F-9712(B). APEX is
a collaboration between the Max-Planck-Institut fur Radioastronomie,
ESO, and the Onsala Space Observatory.

\appendix

\section{The best-fit plane}
\label{ap:A}

This appendix includes different views of the plane \lco\---\lrad\ ---
\lir. The black lines in Fig.~\ref{fig:plane} show the best-fit plane,
shown in eq.~\ref{eq:plane}; the red circles are the galaxies in our
final sample.

Using this plane, it is clear that the dispersion shown is tiny
compared with the large range of IR, RC and CO used in our work.

\begin{figure}
\centering
\includegraphics[width=0.45\textwidth]{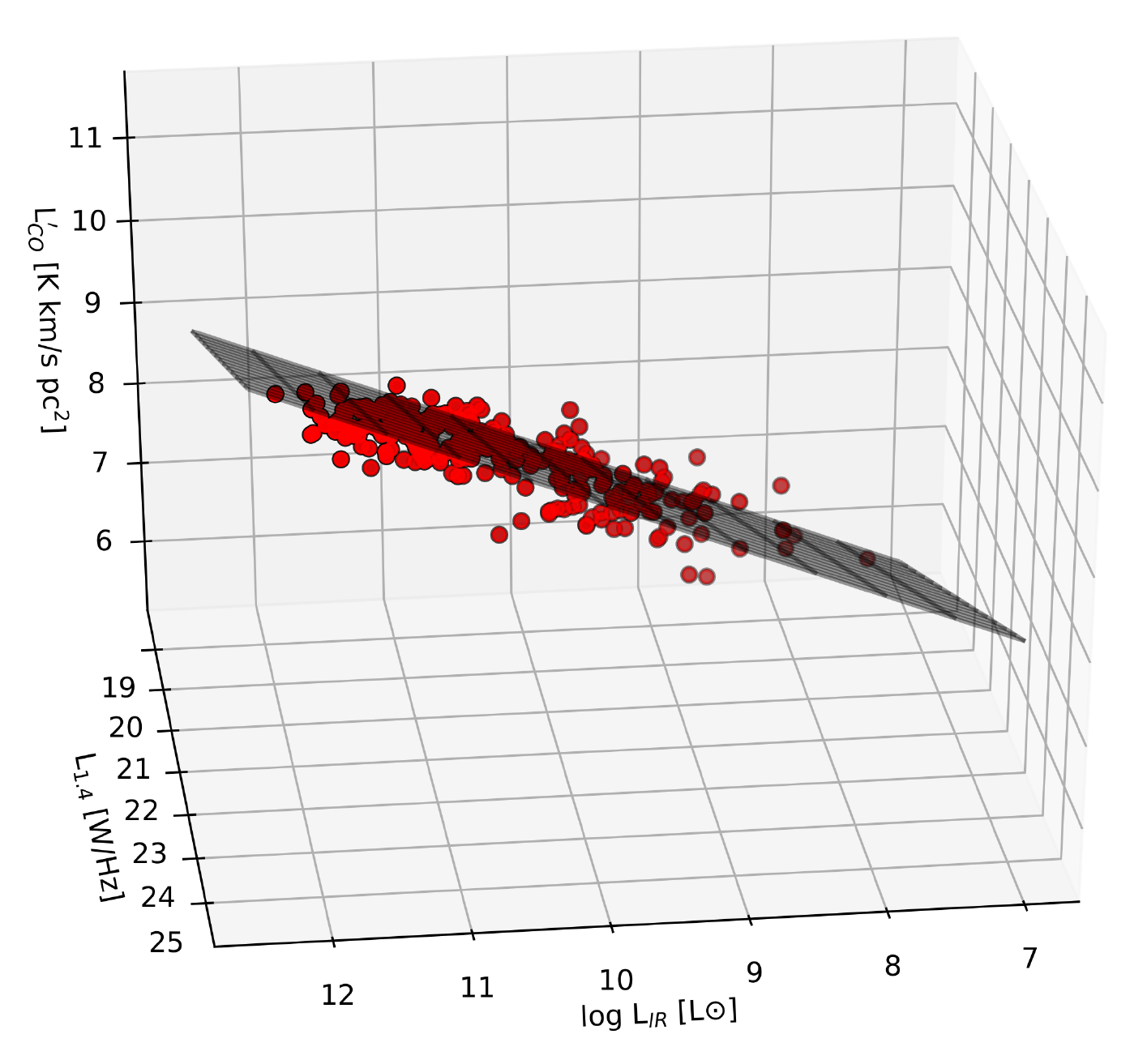}
\includegraphics[width=0.45\textwidth]{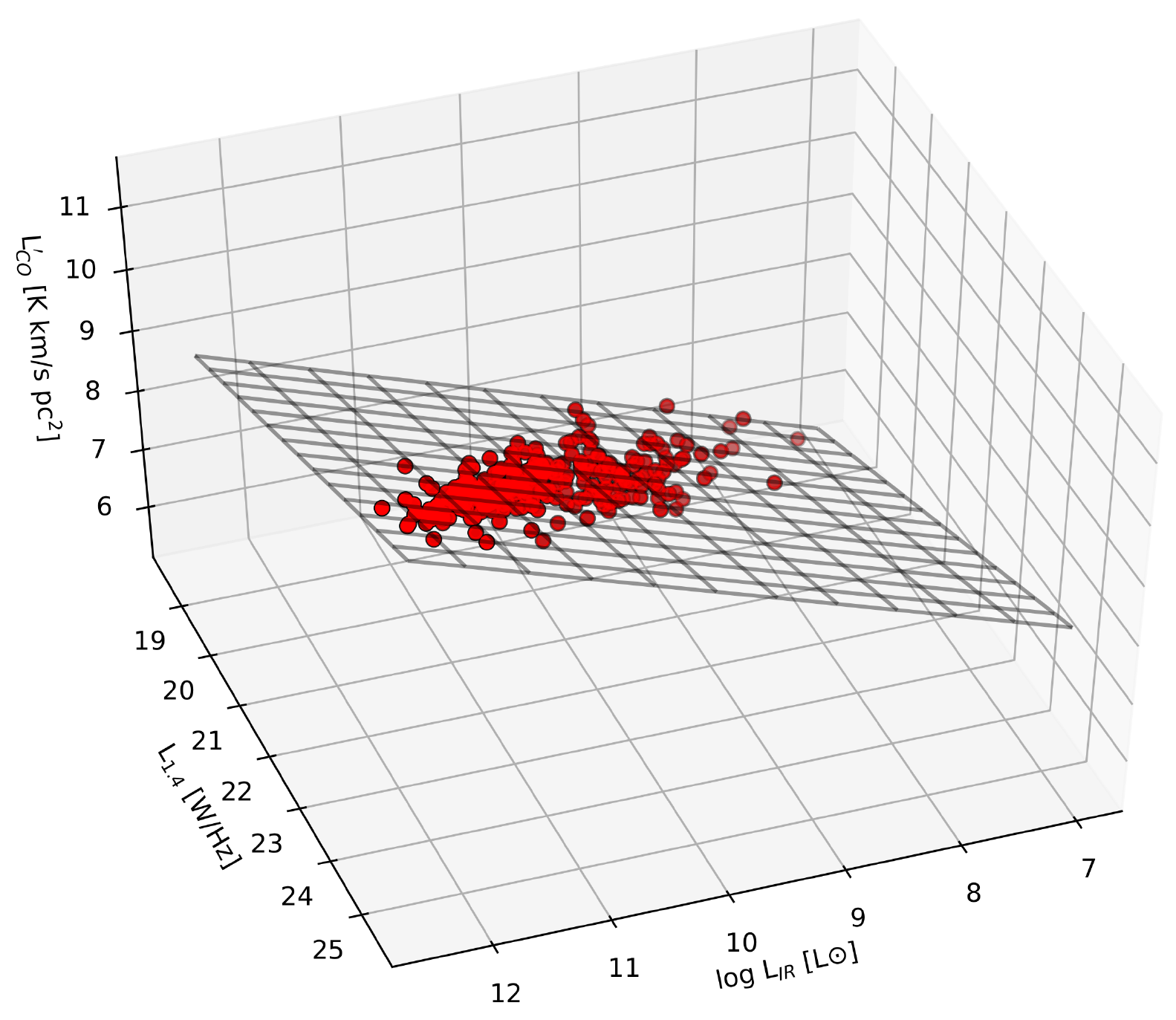}
\includegraphics[width=0.45\textwidth]{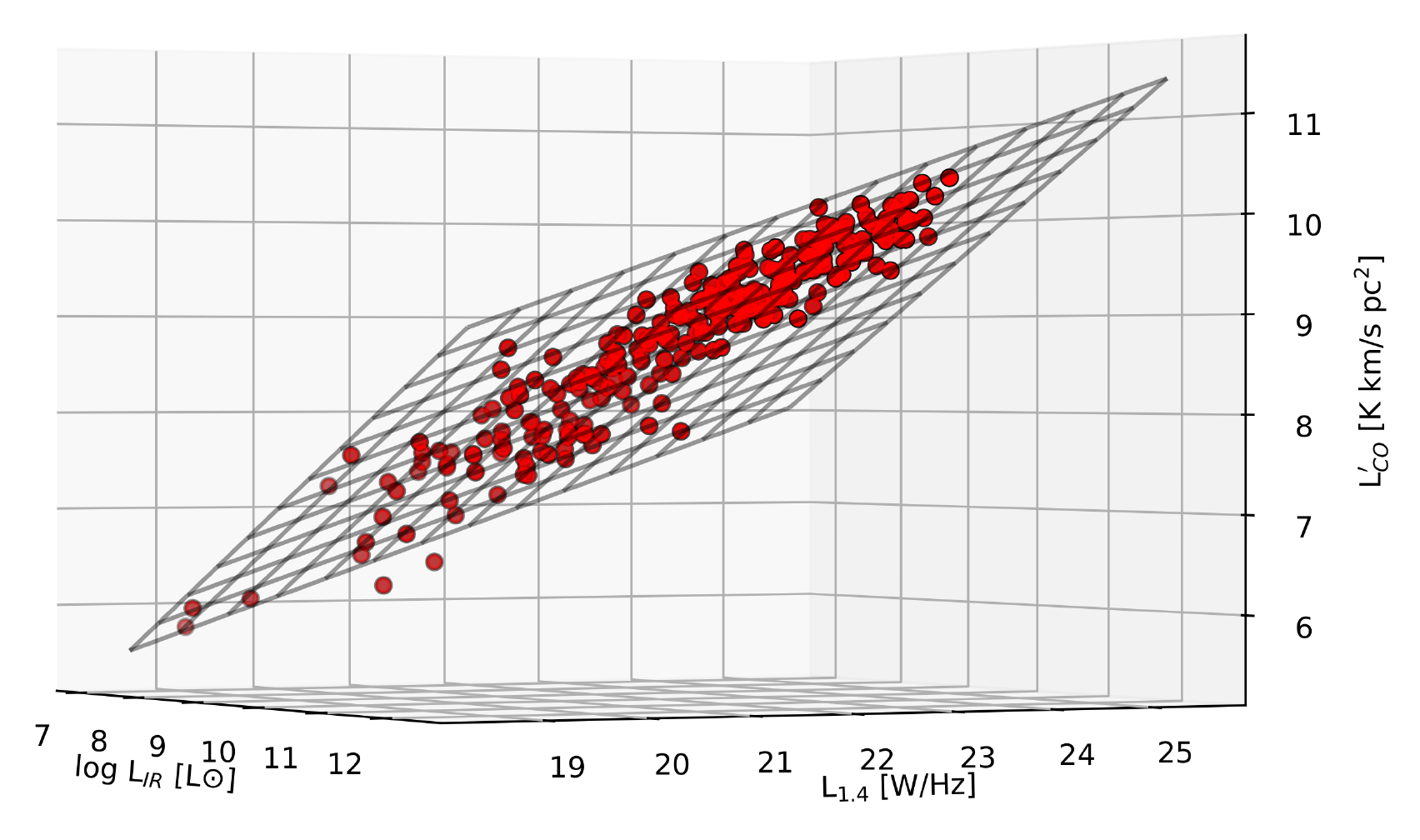}

\caption{Different views of the plane for \lco---\lrad---\lir. The red circles are the
galaxies from the final sample and the black plane is the best fit shown in eq. \ref{eq:plane}.}
\label{fig:plane-final}	
\end{figure}



\bsp	
\label{lastpage}

\begin{thebibliography}{99}
\bibitem[Adler et al.(1991)]{adler91} Adler D.~S., Allen R.~J., Lo K.~Y., 1991, \apj, 382, 475
\bibitem[Amor{\'\i}n et al.(2016)]{amorin16} Amor{\'\i}n, R., Mu{\~n}oz-Tu{\~n}{\'o}n, C., Aguerri, J.~A.~L., et al.\ 2016, \aap, 588, A23
\bibitem[Arimoto et al.(1996)]{arimoto96} Arimoto, N., Sofue, Y., \& Tsujimoto, T.\ 1996, \pasj, 48, 275
\bibitem[Barone et al.(2000)]{barone00} Barone, L.~T., Heithausen, A., H{\"u}ttemeister, S., et al.\ 2000, \mnras, 317, 649
\bibitem[Behroozi et al.(2013)]{behroozi13} Behroozi, P.~S., Wechsler, R.~H., et al.\  2013, \apj, 770, 57 
\bibitem[Bell(2003)]{bell03} Bell, E.~F.\ 2003, \apj, 586, 794
\bibitem[Bianchi(2013)]{bianchi13} Bianchi, S.\ 2013, \aap, 552, A89
\bibitem[Bigiel et al.(2008)]{bigiel08} Bigiel, F., Leroy, A., Walter, F., et al.\ 2008, \aj, 136, 2846
\bibitem[Bolatto et al.(2013)]{bolatto13} Bolatto, A.~D., Wolfire, M., et al.\ 2013, \araa, 51, 207 
\bibitem[Boselli et al.(2002)]{boselli02} Boselli, A., Lequeux, J., \& Gavazzi, G.\ 2002, \apss, 281, 127
\bibitem[Bothwell et al.(2013)]{bothwell13} Bothwell, M.~S., Smail, I., et al.\ 2013, \mnras, 429, 3047 
\bibitem[Cappellari et al.(2011)]{cappellari11} Cappellari, M., Emsellem, E., et al.\ 2011, \mnras, 413, 813 
\bibitem[Carilli $\&$ Walter (2013)]{carilli13} Carilli C.~L., Walter F., 2013, ARA$\&$A, 51, 105
\bibitem[Casasola et al.(2020)]{casasola20} Casasola, V., Bianchi, S., De Vis, P., et al.\ 2020, \aap, 633, A100
\bibitem[Casasola et al.(2017)]{casasola17} Casasola, V., Cassar{\`a}, L.~P., 
Bianchi, S., et al.\ 2017, \aap, 605, A18
\bibitem[Casasola et al.(2015)]{casasola15} Casasola, V., Hunt, L., Combes, F., et al.\ 2015, \aap, 577, A135
\bibitem[Calzetti et al.(2007)]{calzetti07} Calzetti, D., Kennicutt, R.~C., Engelbracht, C.~W., et al.\ 2007, \apj, 666, 870
\bibitem[Cheng et al.(2018)]{cheng18} Cheng, C., Ibar, E., et al.\ 2018, \mnras, 475, 248 
\bibitem[Chung et al.(2009)]{chung09} Chung, A., van Gorkom, J.~H., Kenney, J.~D.~P., et al.\ 2009, \aj, 138, 1741
\bibitem[Cicone et al.(2017)]{cicone17} Cicone, C., Bothwell, M., Wagg, J., et al.\ 2017, \aap, 604, A53
\bibitem[Clark et al.(2015)]{clark15} Clark, C.~J.~R., Dunne, L., Gomez, H.~L., et al.\ 2015, \mnras, 452, 397
\bibitem[Clemens et al.(2013)]{clemens13} Clemens, M.~S., Negrello, M., De Zotti, G., et al.\ 2013, \mnras, 433, 695
\bibitem[Cluver et al.(2014)]{cluver14} Cluver, M.~E., Jarrett, T.~H., et al.\ 2014, \apj, 782, 90 
\bibitem[Condon et al.(1998)]{condon98} Condon, J. J., Cotton, W. D., et al.\ 1998, \aj, 115, 1693
\bibitem[Condon(1992)]{condon92} Condon, J.~J.\ 1992, \araa, 30, 575
\bibitem[Cram et al.(1998)]{cram98} Cram, L., Hopkins, A., Mobasher, B., et al.\ 1998, \apj, 507, 155
\bibitem[Daddi et al.(2010)]{daddi10} Daddi, E., Elbaz, D., et al.\ 2010, \apjl, 714, L118 
\bibitem[Davies et al.(2017)]{davies17} Davies, J.~I., Baes, M., Bianchi, S., et al.\ 2017, \pasp, 129, 044102
\bibitem[Davies et al.(2019)]{davies19} Davies, L.~J.~M., Lagos, C. del P., Katsianis, A., et al.\ 2019, \mnras, 483, 1881
\bibitem[Decarli et al.(2016)]{decarli16} Decarli, R., Walter, F., et al.\ 2016, \apj, 833, 69 
\bibitem[Devereux \& Young(1990)]{devereux90} Devereux, N.~A., \& Young, J.~S.\ 1990, \apj, 359, 42
\bibitem[Devereux, \& Young(1991)]{devereux91} Devereux, N.~A., \& Young, J.~S.\ 1991, \apj, 371, 515
\bibitem[Draine et al.(2007)]{draine07} Draine, B.~T., Dale, D.~A., Bendo, G., et al.\ 2007, \apj, 663, 866
\bibitem[Driver et al.(2016)]{driver16} Driver, S. P., Wright, A. H., et al.\ 2016, \mnras,455, 3911 
\bibitem[Dumas et al.(2011)]{dumas11} Dumas, G., Schinnerer, E., Tabatabaei, F.~S., et al.\ 2011, \aj, 141, 41
\bibitem[Dunne \& Eales(2001)]{dunne01} Dunne, L., \& Eales, S.~A.\ 2001, \mnras, 327, 697
\bibitem[Dunne et al.(2011)]{dunne11} Dunne, L., Gomez, H.~L., da Cunha, E., et al.\ 2011, \mnras, 417, 1510
\bibitem[Eales et al.(2010)]{eales10} Eales, S., Dunne, L., Clements, D., et al.\ 2010, \pasp, 122, 499 
\bibitem[Filho et al.(2019)]{filho19} Filho, M.~E., Tabatabaei, F.~S., S{\'a}nchez Almeida, J., et al.\ 2019, \mnras, 484, 543
\bibitem[Ford et al.(2013)]{ford13} Ford, G.~P., Gear, W.~K., Smith, M.~W.~L., et al.\ 2013, \apj, 769, 55
\bibitem[Freundlich et al.(2019)]{freundlich19} Freundlich, J., Combes, F., Tacconi, L.~J., et al.\ 2019, \aap, 622, A105
\bibitem[Gao \& Solomon(2004)]{gao04} Gao, Y., \& Solomon, P.~M.\ 2004, \apj, 606, 271
\bibitem[Genzel et al.(2010)]{genzel10} Genzel, R., Tacconi, L.~J., et al.\ 2010, \mnras, 407, 2091 
\bibitem[Genzel et al.(2015)]{genzel15} Genzel, R., Tacconi, L.~J., Lutz, D., et al.\ 2015, \apj, 800, 20 
\bibitem[Helfer et al.(2003)]{helfer03} Helfer, T.~T., Thornley, M.~D., Regan, M.~W., et al.\ 2003, \apjs, 145, 259
\bibitem[Helou, \& Bicay(1993)]{helou93} Helou, G., \& Bicay, M.~D.\ 1993, \apj, 415, 93
\bibitem[Helou et al.(1985)]{helou85} Helou, G., Soifer, B.~T., \& Rowan-Robinson, M.\ 1985, \apjl, 298, L7
\bibitem[Hughes et al.(2017a)]{hughes17a} Hughes, T.~M., Ibar, E., et al.\ 2017, \aap, 602, A49 
\bibitem[Hughes et al.(2017b)]{hughes17b} Hughes, T.~M., Ibar, E., Villanueva, V., et al.\ 2017, \mnras, 468, L103
\bibitem[Hunt et al.(2015)]{hunt15} Hunt, L.~K., Garc{\'\i}a-Burillo, S., Casasola, V., et al.\ 2015, \aap, 583, A114
\bibitem[Ibar et al.(2008)]{ibar08} Ibar, E., Cirasuolo, M., Ivison, R., et al.\ 2008, \mnras, 386, 953 
\bibitem[Ibar et al.(2009)]{ibar09} Ibar, E., Ivison, R.~J., et al.\ 2009, \mnras, 397, 281 
\bibitem[Israel, \& Rowan-Robinson(1984)]{israel84} Israel, F., \& Rowan-Robinson, M.\ 1984, \apj, 283, 81
\bibitem[Israel(2000)]{israel00} Israel, F.\ 2000, Molecular Hydrogen in Space, 293
\bibitem[Ivison et al.(2010)]{ivison10} Ivison, R.~J., Magnelli, B., Ibar, E., et al.\ 2010, \aap, 518, L31 
\bibitem[Ivison et al.(2011)]{ivison11} Ivison, R.~J., Papadopoulos,  P.~P., Smail, I., et al.\ 2011, \mnras, 412, 1913
  \bibitem[Jarrett et al.(2011)]{jarrett11} Jarrett, T.~H., Cohen, M., Masci, F., et al.\ 2011, \apj, 735, 112 
\bibitem[Kennicutt \& Evans(2012)]{kennicutt12} Kennicutt, R.~C., \& Evans, N.~J.\ 2012, \araa, 50, 531 
\bibitem[Kennicutt et al.(2011)]{kennicutt11} Kennicutt, R.~C., Calzetti, D., Aniano, G., et al.\ 2011, \pasp, 123, 1347
\bibitem[Kennicutt (1998)]{kennicutt98} Kennicutt R.~C., 1998, ApJ, 498, 541
\bibitem[Kennicutt et al.(2007)]{kennicutt07} Kennicutt, R.~C., Calzetti, D., Walter, F., et al.\ 2007, \apj, 671, 333
\bibitem[Kuno et al.(2007)]{kuno07} Kuno, N., Sato, N., Nakanishi, H., et al.\ 2007, \pasj, 59, 117
\bibitem[Lacki et al.(2010)]{lacki10} Lacki, B.~C., Thompson, T.~A., \& Quataert, E.\ 2010, \apj, 717, 1
\bibitem[Lagos et al.(2018)]{lagos18} Lagos, C.~d.~P., Tobar, R.~J., et al.\ 2018, \mnras, 481, 3573 
\bibitem[Leroy et al.(2005)]{leroy05} Leroy A., Bolatto A.~D., Simon J.~D., Blitz L., 2005, \apj, 625, 763
\bibitem[Lisenfeld et al.(1996)]{lisenfeld96} Lisenfeld, U., Voelk, H.~J., \& Xu, C.\ 1996, \aap, 306, 677
\bibitem[Liu et al.(2015)]{liu15} Liu, L., Gao, Y., \& Greve, T.~R.\ 2015, \apj, 805, 31 
\bibitem[Liu \& Gao(2010)]{liu10} Liu, F., \& Gao, Y.\ 2010, \apj, 713, 524
\bibitem[Madau \& Dickinson (2014)]{madau14} Madau, P., \& Dickinson, M.\ 2014, \araa, 52, 415 
\bibitem[Madau et al.(1996)]{madau96} Madau, P., Ferguson, H.~C., Dickinson, M.~E., et al.\ 1996, \mnras, 283, 1388
\bibitem[Magnelli et al.(2015)]{magnelli15} Magnelli, B., Ivison, R.~J., Lutz, D., et al.\ 2015, \aap, 573, A45 
\bibitem[Magrini et al.(2011)]{magrini11} Magrini, L., Bianchi, S., Corbelli, E., et al.\ 2011, \aap, 535, A13
\bibitem[Mauch et al.(2007)]{mauch07} Mauch, T., \& Sadler, E.~M.\ 2007, \mnras, 375, 931 
\bibitem[Markwardt (2009)]{markwardt09} Markwardt, C.~B.\ 2009, Astronomical Data Analysis Software and Systems XVIII, 411, 251 
\bibitem[McMullin et al.(2007)]{mcmullin07} McMullin, J.~P., Waters, B., et al.\ 2007, Astronomical Data Analysis Software and Systems XVI, 376, 127
\bibitem[Molina et al.(2019)]{molina19} Molina, J., Ibar, E., Villanueva, V., et al.\ 2019, \mnras, 482, 1499
\bibitem[Momose et al.(2013)]{momose13} Momose, R., Koda, J., Kennicutt, R.~C., et al.\ 2013, \apjl, 772, L13
\bibitem[Moshir \& et al.(1990)]{moshir90} Moshir, M., \& et al.\ 1990, IRAS Faint Source Catalogue (ver. 2.0; Pasadena: IPAC)
\bibitem[Murgia et al.(2005)]{murgia05} Murgia M., Helfer T.~T., Ekers R., Blitz L., Moscadelli L., Wong T., Paladino R., 2005, \aap, 437, 389
\bibitem[Murgia et al.(2002)]{murgia02} Murgia M., Crapsi A., Moscadelli L., Gregorini L., 2002, \aap, 385, 412
\bibitem[Neugebauer et al.(1984)]{neugebauer84} Neugebauer, G., Habing, H.~J., et al.\ 1984, \apjl, 278, L1
\bibitem[Niklas, \& Beck(1997)]{niklas97} Niklas, S., \& Beck, R.\ 1997, \aap, 320, 54
\bibitem[Novak et al.(2017)]{novak17} Novak, M., Smol{\v c}i{\'c}, V., Delhaize, J., et al.\ 2017, \aap, 602, A5
\bibitem[Orellana et al.(2017)]{orellana17} Orellana, G., Nagar, N.~M., et al.\ 2017, \aap, 602, A68 
\bibitem[Padovani et al.(2011)]{padovani11} Padovani, P., Miller, N., et al.\ 2011, \apj, 740, 20
\bibitem[Paladino et al.(2006)]{paladino06} Paladino, R., Murgia, M., Helfer, T.~T., et al.\ 2006, \aap, 456, 847
\bibitem[Papadopoulos et al.(2012)]{papadopolous12} Papadopoulos, P.~P., et al.\ 2012, \mnras, 426, 2601 
\bibitem[Popping et al.(2012)]{popping12} Popping, G., Caputi, K.~I., et al. \ 2012, \mnras, 425, 2386 
\bibitem[Pracy et al.(2016)]{pracy16} Pracy, M.~B., Ching, J.~H.~Y., et al.\ 2016, \mnras, 460, 2
\bibitem[Price \& Duric(1992)]{price92} Price, R., \& Duric, N.\ 1992, \apj, 401, 81
\bibitem[Rahman et al.(2011)]{rahman11} Rahman, N., Bolatto, A.~D., Wong, T., et al.\ 2011, \apj, 730, 72
\bibitem[Rickard et al.(1977)]{rickard77} Rickard, L.~J., Palmer, P., Morris, M., et al.\ 1977, \apj, 213, 673
\bibitem[Sadler et al.(2002)]{sadler02} Sadler, E.~M., Jackson, C.~A., Cannon, R.~D., et al.\ 2002, \mnras, 329, 227
\bibitem[Sanders et al.(2003)]{sanders03} Sanders, D.~B., Mazzarella, J.~M., et al. \ 2003, \aj, 126, 1607 
\bibitem[Santini et al.(2014)]{santini14} Santini, P., Maiolino, R., Magnelli, B., et al.\ 2014, \aap, 562, A30
\bibitem[Saintonge et al.(2017)]{saintonge17} Saintonge, A., et al.\ 2017, \apjs, 233, 22 
\bibitem[Saintonge et al.(2011a)]{saintonge11a} Saintonge, A., Kauffmann, G.,  et al.\ 2011, \mnras, 415, 32 
\bibitem[Saintonge et al.(2011b)]{saintonge11b} Saintonge, A., Kauffmann, G.,  et al.\ 2011, \mnras, 415, 61 
\bibitem[Schinnerer et al.(2013)]{schinnerer13} Schinnerer, E., Meidt, S.~E., Pety, J., et al.\ 2013, \apj, 779, 42
\bibitem[Schruba et al.(2012)]{schruba12} Schruba, A., Leroy, A.~K., Walter, F., et al.\ 2012, \aj, 143, 138
\bibitem[Scoville et al.(2016)]{scoville16} Scoville, N., Sheth, K., et al.\ 2016, \apj, 820, 83 
\bibitem[Scoville et al.(2014)]{scoville14} Scoville, N., Aussel, H., Sheth, K., et al.\ 2014, \apj, 783, 84
\bibitem[Semenov et al.(2019)]{semenov19} Semenov, V.~A., Kravtsov, A.~V., \& Gnedin, N.~Y.\ 2019, \apj, 870, 79
\bibitem[Semenov et al.(2017)]{semenov17} Semenov, V.~A., Kravtsov, A.~V., \& Gnedin, N.~Y.\ 2017, \apj, 845, 133
\bibitem[Smith et al.(2014)]{smith14} Smith, D.~J.~B., Jarvis, M.~J., Hardcastle, M.~J., et al.\ 2014, \mnras, 445, 2232
\bibitem[Smol{\v c}i{\'c} et al.(2009)]{smolcic09} Smol{\v c}i{\'c}, V., Schinnerer, E., et al.\ 2009, \apj, 690, 610
\bibitem[Sofue et al.(2003)]{sofue03} Sofue, Y., Koda, J., Nakanishi, H., et al.\ 2003, \pasj, 55, 17
\bibitem[Suchkov et al.(1993)]{suchkov93} Suchkov, A., Allen, R.~J., \& Heckman, T.~M.\ 1993, \apj, 413, 542
\bibitem[Tabatabaei et al.(2013a)]{tabatabei13a} Tabatabaei, F.~S., Schinnerer, E., Murphy, E.~J., et al.\ 2013, \aap, 552, A19
\bibitem[Tabatabaei et al.(2013b)]{tabatabei13b} Tabatabaei, F.~S., Berkhuijsen, E.~M., Frick, P., et al.\ 2013, \aap, 557, A129
\bibitem[Tacconi et al.(2018)]{tacconi18} Tacconi, L.~J., Genzel, R., Saintonge, A., et al.\ 2018, \apj, 853, 179
\bibitem[Tacconi et al.(2013)]{tacconi13} Tacconi, L.~J., Neri, R., Genzel, R., et al.\ 2013, \apj, 768, 74
\bibitem[Thomson et al.(2014)]{thomson14} Thomson, A.~P., Ivison, R.~J., Simpson, J.~M., et al.\ 2014, \mnras, 442, 577
\bibitem[Tinsley \& Danly(1980)]{tisley80} Tinsley, B.~M., \& Danly, L.\ 1980, \apj, 242, 435
\bibitem[V{\'e}ron-Cetty \& V{\'e}ron(2010)]{veron10} V{\'e}ron-Cetty, M.-P., \& V{\'e}ron, P.\ 2010, \aap, 518, A10 
\bibitem[Viaene et al.(2014)]{viaene14} Viaene, S., Fritz, J., Baes, M., et al.\ 2014, \aap, 567, A71
\bibitem[Villanueva et al.(2017)]{villanueva17} Villanueva, V., Ibar, E., et al.\ 2017, \mnras, 470, 3775 
\bibitem[Voelk(1989)]{voelk89} Voelk, H.~J.\ 1989, \aap, 218, 67
\bibitem[Walter et al.(2016)]{walter16} Walter, F., Decarli, et al.\ 2016, \apj, 833, 67 
\bibitem[Wenger et al.(2000)]{wenger00} Wenger, M., Ochsenbein, F., Egret, D., et al.\ 2000, \aaps, 143, 9
\bibitem[White et al.(1998)]{white98} White, R. L., Becker, R. H., et al.\ 1998, \apj, 475, 47
\bibitem[Wilson(1995)]{wilson95} Wilson, C.~D.\ 1995, \apjl, 448, L97
\bibitem[Wright et al.(2010)]{wright10} Wright, E.~L., Eisenhardt, P.~R.~M., Mainzer, A.~K., et al.\ 2010, \aj, 140, 1868
\bibitem[Xu, \& Helou(1996)]{xu96} Xu, C., \& Helou, G.\ 1996, \apj, 456, 163
\bibitem[Young et al.(2011)]{young11} Young, L.~M., Bureau, M., et al.\ 2011, \mnras, 414, 940 
\bibitem[Young et al.(2008)]{young08} Young, L.~M., Bureau, M., \& Cappellari, M.\ 2008, \apj, 676, 317
\bibitem[Young, \& Scoville(1991)]{young91} Young, J.~S., \& Scoville, N.~Z.\ 1991, \araa, 29, 581
\bibitem[Yun et al.(2001)]{yun01} Yun, M.~S., Reddy, N.~A., Condon, J.~J.\ 2001, \apj, 554, 803
\end{thebibliography}
\end{document}